\documentclass{aastex62}

\usepackage{graphicx}	
\usepackage{amssymb}
\usepackage{amsmath}
\usepackage{natbib}
\usepackage{color}
\usepackage{url}
\usepackage{placeins}
\usepackage{rotating}

\newcommand{\ie}{{\it i.e.\,}}
\newcommand{\eg}{{\it e.g.}}
\newcommand{\insitu}{{\it in situ~}}
\newcommand{\kmps}{{$\mathrm{~km~s}^{-1}$}}

\newcommand{\gammaunit}{{$\times10^{-7} ~\mathrm{km}^{-1}$\,}}
\newcommand{\Rsun}{{$~\mathrm{R}_{\odot}$}}
\newcommand{\angstrom}{\textup{\AA}}

\submitjournal{ApJ}

\shorttitle{July 2017 event}
\shortauthors{Dumbovic et al.}

\begin{document}

\title{Unusual plasma and particle signatures at Mars and STEREO-A related to CME-CME interaction}

\correspondingauthor{Mateja Dumbovi\'{c}}
\email{mateja.dumbovic@uni-graz.at}

\correspondingauthor{Jingnan Guo}
\email{jnguo@ustc.edu.cn}

\author[0000-0002-8680-8267]{Mateja Dumbovi\'{c}}
\affil{Institute of Physics, University of Graz, Universit\"atsplatz 5, A-8010 Graz, Austria}

\author[0000-0002-8707-076X]{Jingnan Guo}
\affiliation{CAS Key Laboratory of Geospace Environment, School of Earth and Space Sciences, University of Science and Technology of China, Hefei, China}
\affiliation{Department of Extraterrestrial Physics, Christian-Albrechts University in Kiel, Leibnitzstrasse 11, 24098, Kiel, Germany}

\author[0000-0003-4867-7558]{Manuela Temmer}
\affil{Institute of Physics, University of Graz, Universit\"atsplatz 5, A-8010 Graz, Austria}

\author[0000-0001-9177-8405]{M. Leila Mays}
\affiliation{NASA Goddard Space Flight Center, Greenbelt, MD 20771, USA}

\author[0000-0003-2073-002X]{Astrid Veronig}
\affil{Institute of Physics, University of Graz, Universit\"atsplatz 5, A-8010 Graz, Austria}
\affil{Kanzelh\"ohe Observatory for Solar and Environmental Research, University of Graz, Kanzelh\"ohe 19, 9521 Treffen, Austria}

\author[0000-0002-2655-2108]{Stephan Heinemann}
\affiliation{Institute of Physics, University of Graz, Universit\"atsplatz 5, A-8010 Graz, Austria}

\author[0000-0001-5661-9759]{Karin Dissauer}
\affiliation{Institute of Physics, University of Graz, Universit\"atsplatz 5, A-8010 Graz, Austria}

\author[0000-0001-7662-1960]{Stefan Hofmeister}
\affiliation{Institute of Physics, University of Graz, Universit\"atsplatz 5, A-8010 Graz, Austria}

\author[0000-0001-5258-6128]{Jasper Halekas}
\affiliation{Department of Physics and Astronomy, University of Iowa, Iowa City, IA 52242, USA}

\author[0000-0001-6868-4152]{Christian M\"ostl}
\affiliation{Space Research Institute, Austrian Academy of Sciences, Schmiedlstraße 6, 8042 Graz, Austria}

\author[0000-0001-9024-6706]{Tanja Amerstorfer}
\affiliation{Space Research Institute, Austrian Academy of Sciences, Schmiedlstraße 6, 8042 Graz, Austria}

\author[0000-0002-1222-8243]{J\"urgen Hinterreiter}
\affiliation{Space Research Institute, Austrian Academy of Sciences, Schmiedlstraße 6, 8042 Graz, Austria}
\affiliation{Institute of Physics, University of Graz, Universit\"atsplatz 5, A-8010 Graz, Austria}

\author[0000-0003-0724-9063]{Sa\v{s}a Banjac}
\affiliation{Department of Extraterrestrial Physics, Christian-Albrechts University in Kiel, Leibnitzstrasse 11, 24098, Kiel, Germany}

\author[0000-0001-5622-4829]{Konstantin Herbst}
\affiliation{Department of Extraterrestrial Physics, Christian-Albrechts University in Kiel, Leibnitzstrasse 11, 24098, Kiel, Germany}

\author[0000-0002-8887-3919]{Yuming Wang}
\affiliation{CAS Key Laboratory of Geospace Environment, School of Earth and Space Sciences, University of Science and Technology of China, Hefei, China}

\author[0000-0003-4992-4725]{Lukas Holzknecht}
\affiliation{Institute of Physics, University of Graz, Universit\"atsplatz 5, A-8010 Graz, Austria}

\author[0000-0002-3776-2776]{Martin Leitner}
\affiliation{Institute of Physics, University of Graz, Universit\"atsplatz 5, A-8010 Graz, Austria}

\author[0000-0002-7388-173X]{Robert F. Wimmer--Schweingruber}
\affiliation{Department of Extraterrestrial Physics, Christian-Albrechts University in Kiel, Leibnitzstrasse 11, 24098, Kiel, Germany}

\begin{abstract}
On July 25 2017 a multi-step Forbush decrease (FD) with the remarkable total amplitude of more than 15\% was observed by MSL/RAD at Mars. We find that these particle signatures are related to very pronounced plasma and magnetic field signatures detected in situ by STEREO-A on July 24 2017, with a higher than average total magnetic field strength reaching more than 60 nT. In the observed time period STEREO-A was at a relatively small longitudinal separation (46 degrees) to Mars and both were located at the back side of the Sun as viewed from Earth. We analyse a number of multi-spacecraft and multi-instrument (both in situ and remote-sensing) observations, and employ modelling to understand these signatures. We find that the solar sources are two CMEs which erupted on July 23 2017 from the same source region on the back side of the Sun as viewed from Earth. Moreover, we find that the two CMEs interact non-uniformly, inhibiting the expansion of one of the CMEs in STEREO-A direction, whereas allowing it to expand more freely in the Mars direction. The interaction of the two CMEs with the ambient solar wind adds up to the complexity of the event, resulting in a long, sub-structured interplanetary disturbance at Mars, where different sub-structures correspond to different steps of the FD, adding-up to a globally large-amplitude FD.
\end{abstract}


\section{Introduction}
\label{intro}
Interplanetary counterparts of coronal mass ejections (ICMEs) are the most prominent short-term transient phenomena in the heliosphere which significantly influence the interplanetary space and can have major impact on Earth and other planets \citep[see an overview by][and references therein]{kilpua17}. The Heliophysics System Observatory\footnote{\url{https://www.nasa.gov/content/goddard/heliophysics-system-observatory-hso}} enables us to observe ICME-related signatures at different heliospheric positions and, combined with remote observations as well as modelling efforts, to better understand the propagation and evolution of ICMEs. Recently, an ICME (\ie its shock signatures) was tracked and modelled all the way from the Sun to the outer heliosphere \citep{witasse17}, using, among other things, Forbush decreases (FDs) as ICME signatures.

FDs are observed as short-term depressions in the galactic cosmic ray flux \citep{forbush37,hess37}, with onset corresponding to the ICME arrival \citep{cane96,dumbovic11}. The magnitude, shape, duration as well as the sub-structuring of the FD depend on the physical properties of the corresponding interplanetary transient \citep[for overview see \eg][and references therein]{richardson04,cane00,belov09} and more specifically, ICME-related FDs are expected to reflect the evolutionary properties of ICMEs, such as expansion \citep{dumbovic18b}. Therefore, they are highly suitable as ICME signatures and are used to indicate ICME arrival \textit{when} and \textit{where} other \insitu measurements are unavailable \citep[\eg\, in pre-satellite era or at Mars, see][]{mostl15,lefevre16,vennerstrom16,forstner17,winslow18}. From that perspective, the Radiation Assessment Detector \citep[RAD,][]{hassler12}, on board Mars Science Laboratory's (MSL) rover Curiosity \citep{grotzinger12} was shown to be highly suitable for identifying ICMEs' arrival at Mars with detected FDs \citep{guo18}.

While trying to understand the basic physics, often single and simple, text--book example ICMEs are analysed, but it is important to note that the frequency of such events can be relatively low. Many \insitu detected ICMEs do not show typical magnetic cloud properties \citep[\ie magnetic field rotation, low plasma beta, low density and temperature, expanding speed profile, see \eg][and references therein]{burlaga81,zurbuchen06,kilpua17}, quite likely because of, among other things, CME-CME interactions. CME-CME interactions are expected to be quite common considering the CME occurrence frequency and their typical propagation time \citep{lugaz17} and can lead to more intense geomagnetic storms \citep{wang03a, wang03b, farrugia06a, farrugia06b, xie06, lugaz14, shen17} as well as larger-amplitude FDs \citep{papaioannou10,dumbovic16}. Therefore, from the space weather point of view these complex events are especially challenging and interesting to study \citep[see also \eg][]{webb17}. Moreover, understanding the underlying physics in such events offers an opportunity to improve future modelling efforts and predictions of such strong space weather events.

\section{Data and method}
\label{data}	

On July 25 2017 MSL/RAD observed a Forbush decrease (FD) with one of the biggest relative amplitudes detected by this instrument since its launch \citep{guo18}. The FD does not only have an unusually large amplitude, but in addition shows a quite complex time profile, possibly indicating interactions of two or even several interplanetary transients. In order to understand this event and the conditions which lead to these specific signatures we search for possible solar sources, assuming that (1) a very impulsive CME is most likely involved, and (2) CME-CME interaction is very likely. 

In a time period 6 days prior to the FD observed at Mars, we search for possible CME candidates. For this we use SOHO/LASCO  \citep{brueckner95} coronagraphs C2 and C3, with field-of-view reaching 6 \Rsun\, and 32 \Rsun, respectively, and STEREO-A(ST-A)/SECCHI \citep{howard08} EUVI/COR1/COR2 image data with field-of-view reaching 1.7 \Rsun, 2.5 \Rsun\, and 15 \Rsun, respectively. We find only 4 CMEs that have a position angle relating to Mars direction, which was almost in the oposition. The first two CMEs launched on July 20, and the other two launched about two days later, around the beginning of July 23 and were much stronger and faster. The last of these 4 CMEs is extremely fast and wide and is very likely interacting with a preceding CME launched shortly before. As the separation angle between ST-A and Mars is not large ($46^{\circ}$), we check possible ICME signatures measured at ST-A for more insight on these ICME characteristics. For this we use the PLASTIC \citep{galvin08} and IMPACT \citep{acuna08} \insitu plasma and magnetic field instruments. Indeed, we find a very pronounced \insitu magnetic cloud signature on July 24 2017 with a magnetic field strength reaching more than 60 nT. 

The two CMEs that erupted on July 23 are the most likely candidates of the signatures observed both in MSL/RAD and ST-A/PLASTIC+IMPACT as they best match the timing and direction for causing the observed FD at Mars. Therefore, in the following we focus on these two CMEs, here forth denoted as CME1 and CME2. The other two CMEs, launched on July 20, are discussed later in the frame of background solar wind effects and are denoted as CMEs 0.1 and 0.2. We hypothesise that the unusual observational signatures are due to the interaction of CMEs 1 and 2. To test this hypothesis we employ a number of multi-spacecraft and multi-instrument observations, as well as modelling. We note that the observational methods and models we use to test the hypothesis suffer from a number of uncertainties, therefore, constructing a plausible explanation of this complex event is challenging. Nevertheless, we argue that the combined synergetic view of the results and interpretations obtained from different models and observations indicates the most likely explanation.

\begin{figure}[ht!]
\plotone{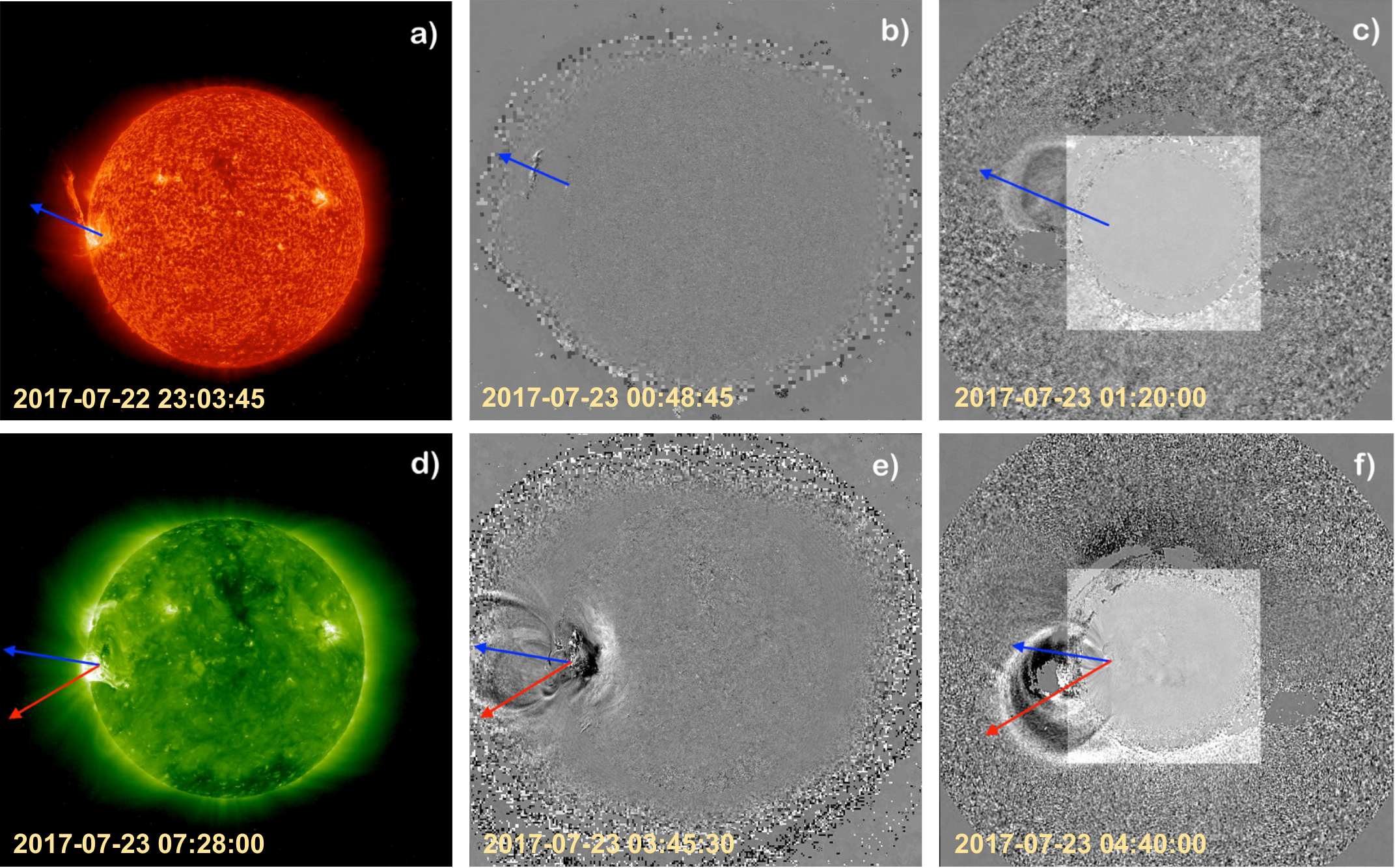}
\caption{Low coronal eruption signatures associated with CMEs 1 and 2:
a) a prominence eruption associated with CME1 observed by ST-A/EUVI 304 \angstrom; 
b) running difference image of ST-A/EUVI 304 \angstrom\, showing prominence moving radially outward;
c) running difference images of ST-A/EUVI 304 \angstrom\, and ST-A/COR1 showing CME1;
d) post-flare loops associated with CME2 observed by ST-A/EUVI 195 \angstrom;
e) running difference image of ST-A/EUVI 195 \angstrom\, with erupting loops associated to CME2;
f) running difference images of ST-A/EUVI 195 \angstrom\, and ST-A/COR1 showing the ``main" part of CME2.
Blue arrow in a-c marks the direction of the CME1 eruption. Red and blue arrows in d-f mark the directions of multi--step eruptions observed in ST-A/EUVI (For detailed explanation see main text. \textit{Credit: JHelioviewer}).
\label{fig1}}
\end{figure}

\subsection{CME observations}
\label{cme}	

A filament/prominence eruption starts around 22:30 UT on July 22 2017 observed by ST-A/EUVI 304 \angstrom\; (Figure \ref{fig1}a), where one part of the filament continues to move radially outward (Figure \ref{fig1}b) and can be observed as CME in COR1 at around 01:05 UT on July 23 2017 with the same (northward) direction of motion (Figure \ref{fig1}c). The source region appears to be AR 12665 at the east limb of ST-A's field of view with approximate Stonyhurst coordinates [0,155]. The CME (henceforth CME1) is observed in COR2 at 01:55 UT, and appears in LASCO/C2 field of view at 01:36 UT.

 \begin{figure*}
\gridline{\fig{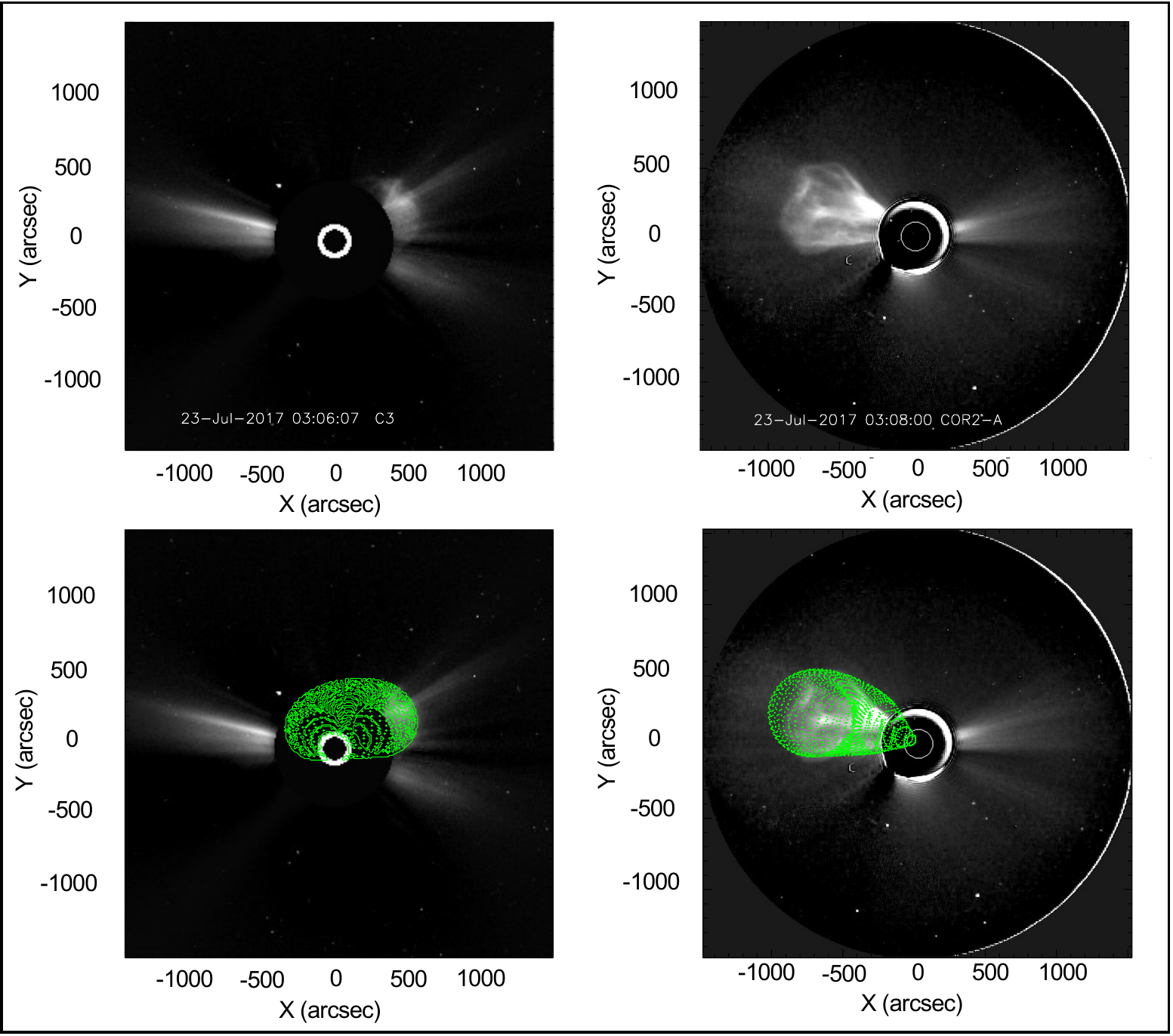}{0.48\textwidth}{(a)}
          \fig{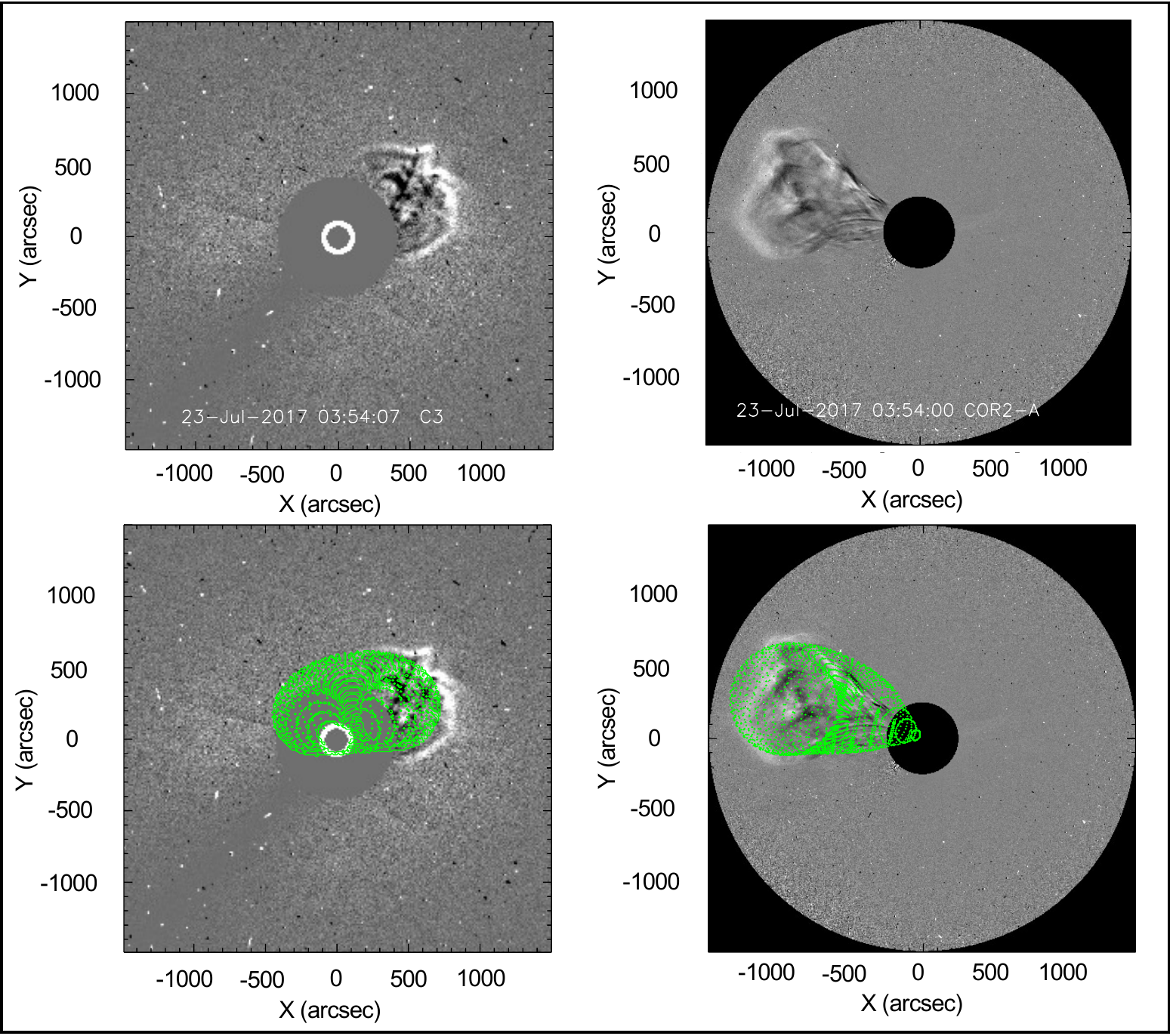}{0.48\textwidth}{(b)}}
\gridline{\fig{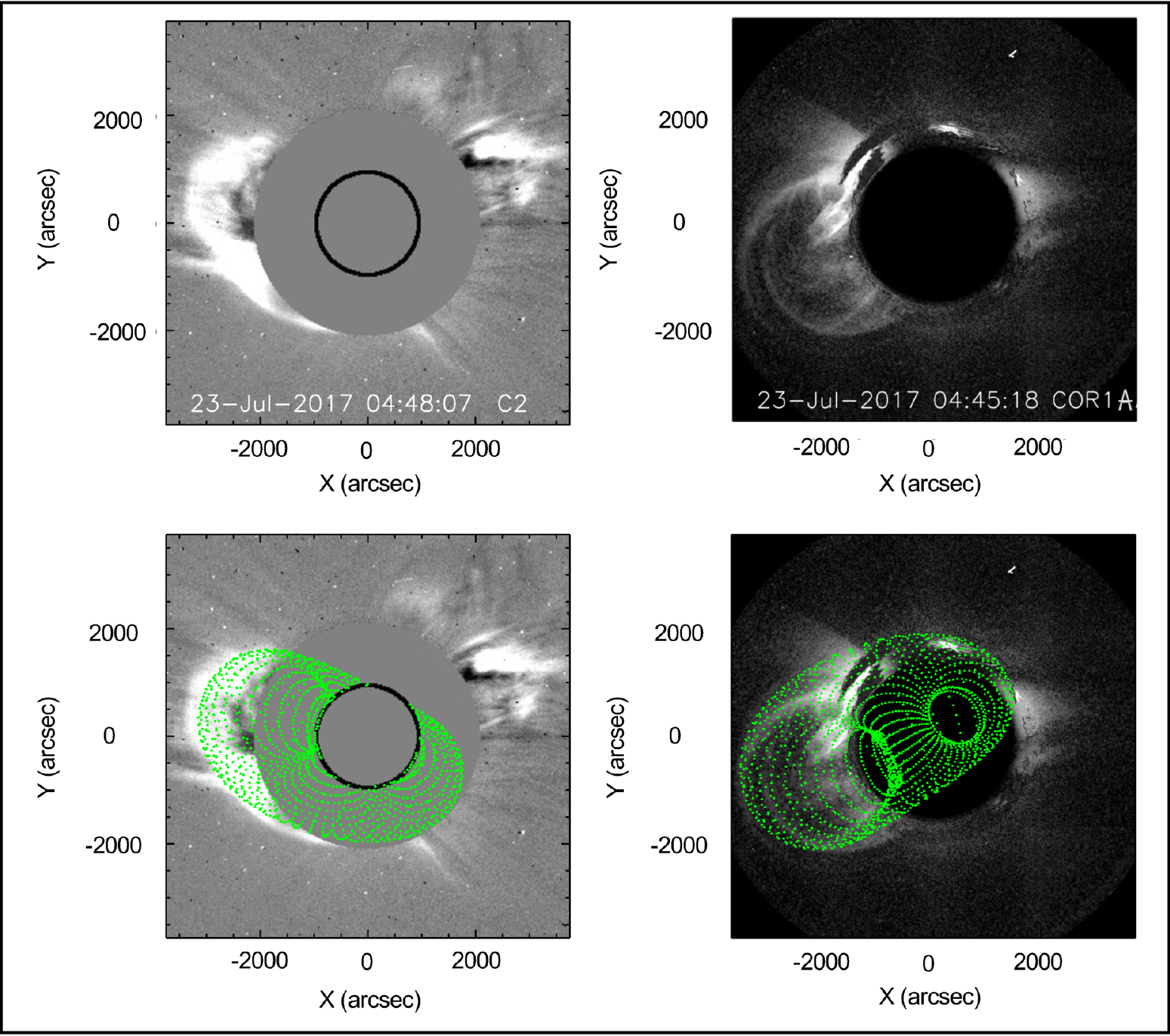}{0.48\textwidth}{(c)}
          \fig{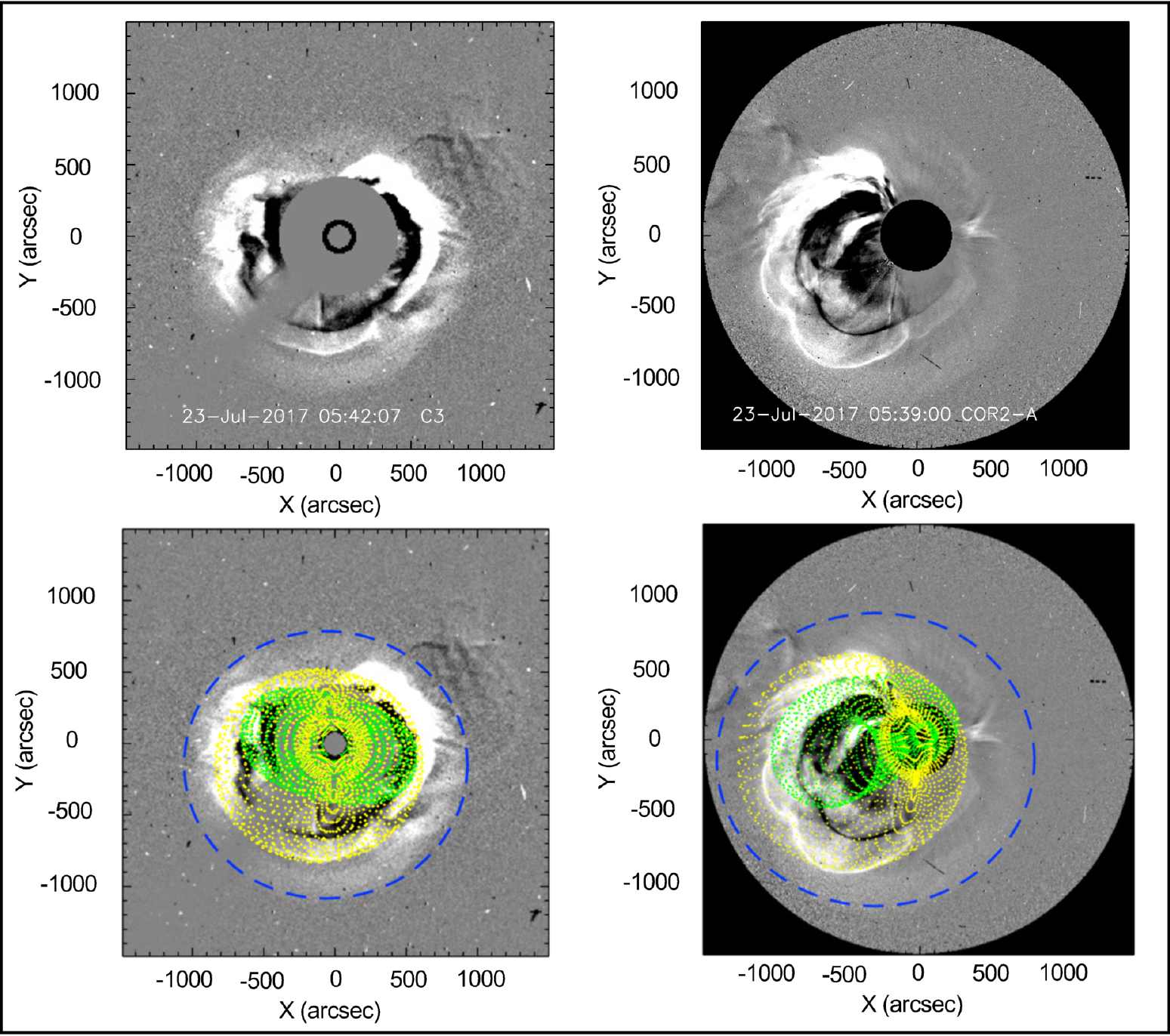}{0.48\textwidth}{(d)}}
\caption{GCS reconstructions of CMEs 1 and 2 at different times:
a) when the CME1 leading edge can be fitted with a croissant ($R=11.6$\Rsun);
b) when the leading edge of CME1 cannot be perfectly fitted anymore($R=14.9$\Rsun);
c) the ``main" part of CME2 fitted with a croissant model (in C2 and COR1, $R=4.25$\Rsun);
d) a multi--step complex CME2 fitted with a quasi-sphere (in C3 and COR2, $R=19.2$\Rsun).
The green mesh outlines the fitted croissant both for CME 1 and 2, the yellow wire outlines a fitted quasi-sphere for CME2 and the blue dashed circle outlines the shock front. Reconstruction parameters for both CMEs are given in Table \ref{tab2}.
\label{fig2}}
\end{figure*}

In order to more reliably derive the direction and geometry of the CME, we use coronagraphic images from different vantage points, ST-A and LASCO, to perform a 3D CME reconstruction using the Graduated cylindrical shell model \citep[GCS,][]{thernisien06,thernisien09,thernisien11}. GCS assumes that geometrically a CME can be represented as a hollow croissant with origin in the center of the Sun, \ie with conical legs, circular cross section and pseudo-circular front. A projection of the croissant (green mesh in Figure \ref{fig2}) at a given time is fitted to coronagraphic images, where two different vantage points are needed to better constrain the fit. We manually fit the GCS model to the CME observations using ST-A/COR2 and LASCO/C3 images from 03:06 and 03:08 UT respectively (see Figure \ref{fig2}a). Assuming radial propagation direction, we use as starting position the CME source region, \ie the AR location. The best fit is derived for a direction slightly different from the AR location (Stonyhurst coordinates [10,170]) indicating a deflection towards Mars and ST-A. For obtaining a kinematic profile, we assume self--similar expansion, keep all GCS parameters, except for height, constant and fit stereoscopic images for several time steps. We note that for later time steps we cannot find a good match anymore for the CME front as it deforms most probably due to a helmet streamer (see Figure \ref{fig2}b).
 
As CME1 exits ST-A/COR2 field of view around 03:20 UT, an eruption is observed in the same AR 12665 at the east limb in ST-A/EUVI 195 \angstrom, followed by a dimming, an EUV wave and post-flare loops (fig \ref{fig1}d). A more detailed analysis of the EUVI observations reveals a multi--step eruption, where several eruptions are observed very closely in time in the same AR. We first observe erupting loops in EUVI, appearing also in COR1 but as very faint structures and are oriented northward (blue arrow in Figure \ref{fig1}d-e). The second eruption observed in EUVI is associated with a strong EUV wave and is seen as a very prominent structure in COR1 moving more southward (red arrow in Figure \ref{fig1}d-e). The third eruption is associated with a strong flare, is narrow, and the most impulsive. In COR1 it is not easily distinguishable from the second eruption, however, it can be seen as a localised emission enhancement, where its outer structure overlap with the Northern loops segment of the previous eruption. For a more detailed analysis of the early multi--eruption signatures we refer the reader to a study by \citet{liu19}. The single eruptions cannot be individually tracked and appear to move as one entity already in COR1 field-of-view. In COR2 field-of-view the structures appear as a single CME. Therefore, in the further analysis we treat this as a multi--step eruption resulting in a single CME. The CME (henceforth CME2) is detected by COR2 at 04:54 UT, and appears in the LASCO/C2 field of view at 04:45 UT. 

We perform a GCS reconstruction of the CME main part in ST-A/COR1 and LASCO/C2 images at 04:45 and 04:48, respectively, where the CME can be reasonably fitted by a croissant model (green mesh in Figure \ref{fig2}c). However, in the field of view of COR2 and C3 the simple croissant geometry does not match anymore with the observed CME structure, and a best fit would be obtained for a quasi--spherical geometry (yellow mesh in Figure \ref{fig2}d)\footnote{We note that it is unlikely that this bright structure is a shock, as the typical shock signature is a faint front \citep{vourlidas13} which we observe propagating ahead (blue dashed circle in Figure \ref{fig2}d)}. Even by introducing deflection, rotation and non--self--similar expansion it is impossible to obtain a reasonable transition from the croissant fitted in COR1 to the quasi--sphere fitted in COR2. This reflects the interaction of these multiple eruptions very early in their evolution. We therefore abandon the ``main" eruption part fitted in COR1/C2 and for obtaining the kinematics of CME2 track the quasi--spherical bright structure, which we extrapolate back and forward in time assuming self-similar expansion and radial propagation.

\subsection{\textit{In situ} observations at STEREO-A}
\label{sta}	

\begin{sidewaysfigure}
\centerline{\includegraphics[width=0.85\textwidth]{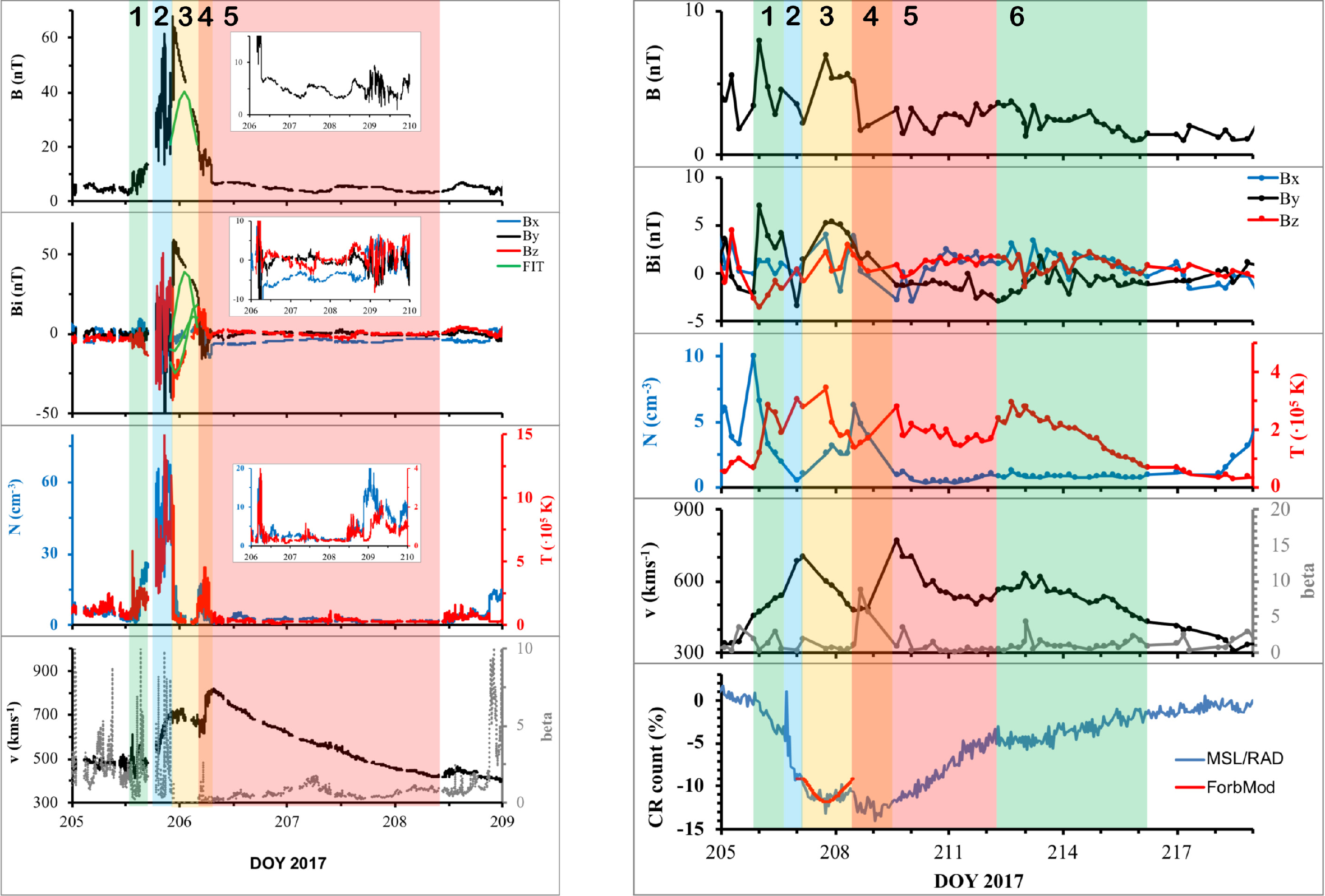}}
\caption{\textit{In situ} measurements at ST-A and Mars:
\newline \textit{Left:} ST-A/PLASTIC+IMPACT measurements for July 24--28 (DOY 205--209) with top-to-bottom panels showing magnetic field strength, $B$, three magnetic field components, $B_x$, $B_y$, and $B_z$, plasma density, $N$, plasma temperature, $T$, and plasma beta and speed, $v$. Differently coloured shading outlines different regions (1--5, for explanation see main text). Small panels embedded in top three larger panels show a zoomed-in region 5 (DOY 206--210). The green curves in top two panels mark the Lundquist fit (hourly values, $\chi^2$-test p-value is 0.011).
\newline \textit{Right:} MAVEN and MSL/RAD measurements for July 24 -- August 7 (DOY 205--219) with top-to-bottom panels showing magnetic field strength, $B$, three magnetic field components, $B_x$, $B_y$, and $B_z$, plasma density, $N$ and temperature, $T$, plasma beta and speed, $v$, and GCR count proxy by MSL/RAD. Differently coloured shading outlines different regions (1--6, for explanation see main text). The red curve in the bottom plot marks the ForbMod fit (Forbush decrease model, for explanation see section \ref{evolution}).
}
\label{fig3}
\end{sidewaysfigure}

In Figure \ref{fig3} ST-A/IMPACT and PLASTIC \insitu magnetic field and plasma data are shown in a time period 24--28 July 2017. The \insitu event shows quite complex signatures, where several different regions can be identified using standard observational criteria described by \eg\, \cite{zurbuchen06,richardson10,kilpua17,richardson18}. The start of the event can be identified on July 24 (DOY 205) around 13:15 UT with a simultaneous jump in magnetic field, temperature, density and speed followed by the typical sheath characteristics - elevated temperature, density, speed and plasma beta and fluctuating magnetic field (region 1 in Figure \ref{fig3}). Although the elevated values do not seem prominent compared to the rest of the interval (extreme parameter values in regions 3 and 4) we note that the magnetic field increases from around 3 nT to around 11 nT, a jump comparable to typical events. Therefore, we identify this region as the shock/sheath region.

In region 2 we also observe typical sheath characteristics, however they are much stronger than in region 1: the magnetic field strength is on average 4--5 times stronger, density is about 6 times higher, temperature 2--3 times higher and there is an increase of flow speed throughout the region from about 550 \kmps to 660 \kmps. This might indicate a second shock/sheath region or a continuation of one common sheath where the plasma and magnetic field are strongly piled up against the strong magnetic field of the magnetic structure in the region that follows (region 3). However, due to the data gap between region 1 and 2, this can not be fully resolved.

In region 3 qualitatively typical magnetic cloud (MC) signatures are observed -- strong and smoothly rotating magnetic field, low density, temperature and plasma beta, however quantitatively this MC is quite unusual. The magnetic field strength is extremely high, at its peak ($\sim65$ nT) it is 6 times higher than MC values typically observed at 1 AU \citep[10 nT at Earth, see][]{richardson10}, the MC duration ($\sim7$ hours) is on the other hand extremely short, around 4 times shorter than MC durations typically observed at 1 AU \citep[around 30 hours, see][]{richardson10}, and finally, the flow speed does not show a typical linearly decreasing (expansion) profile. These are strong indications that the expansion of this CME was inhibited. We perform the Lundquist fitting to the magnetic field data \citep{leitner07} and find a reasonable fit (green curves in Figure \ref{fig3}b) with orientation showing a $-13^{\circ}$ axis tilt to ecliptic, central magnetic field of 40 nT and an estimated diameter of 0.12 AU, \ie 26 \Rsun. There is a quite large discrepancy of the fit and data at the beginning of the MC, as the magnetic field shows a quite asymmetric profile, not unusual for the fast CMEs \citep{masias-meza16}, which might be due to a plasma/magnetic field line pile--up within the MC near the leading edge.

In region 4 we observe a typical stream interface signature -- drop of density followed by the increase in temperature and speed, the fluctuating and elevated magnetic field and small spike in the plasma beta, a common feature of stream interaction regions (SIRs). This is however not followed by the typical high speed stream (HSS) signatures in region 5, where high flow speed is accompanied with a relatively low temperature, plasma beta and density. The speed which peaks on the front with about 800 \kmps shows a highly expanding profile with speed at the trailing edge reaching 400 \kmps, and the duration of the region is about 50h, which is quite long compared to typical ICME duration at 1 AU \citep[around 30 hours, see][]{richardson10}. The magnetic field is relatively smooth but weak and shows no obvious rotation, although there are indications of sub-structuring. These might be indications of a highly expanded CME, a specific trajectory of the spacecraft through the leg of the CME \citep[see \eg][]{mostl10}, or of a complex ejecta/compound stream \citep[see e.g.][]{burlaga03,lugaz17}.

\subsection{\textit{In situ} observations at Mars}
\label{mars}

In Figure \ref{fig3} MAVEN \citep[Mars Atmosphere and Volatile EvolutioN,][]{jakosky15} \insitu magnetic field and plasma data, as well as MSL/RAD relative count rates are shown in a time period 24 July -- 7 August 2017. To avoid contamination from measurements taken inside the bow shock we use an algorithm described by \citet{halekas17}. Similar as at STEREO-A, the \insitu event at Mars shows quite complex signatures, where several different regions can be identified. We note that the identification here is not so straightforward as with ST-A, due to uneven coverage of the solar wind. We interpolate the measurements to guide the eye, however, we note that one has to be careful not to misinterpret the interpolation for measurements. Nevertheless, using MAVEN in combination with MSL/RAD data, which provides additional information on the associated Forbush decreases, we are able to identify different sub-structures. For that purpose we use sol--filtered MSL/RAD data in which the daily pressure--induced oscillations have been removed \citep[for details on the method see][]{guo18}. As a proxy of the CR count we use relative values of the dose rate normalised to the beginning of the observed period.

Region 1 is characterised by, what appears to be typical stream interface properties with a corresponding decrease in GCRs of about 4\% as detected by MSL/RAD which starts July 25 around 00:00 UT. Unfortunately, MAVEN was not measuring the solar wind plasma and magnetic field throughout region 2, however we do discern this as a separate region due to two reasons. Firstly, the plasma speed and the $B_y$ component show distinctively different properties at the start and end of this gap. Secondly, MSL/RAD data show an increase of 5\% at the start of this region, followed by a rapid decrease of 5\% (measured from the onset point, not the peak point) which are typical shock/sheath-related FD properties \citep[\eg][]{cane00}. Therefore, we suggest that this is most likely a shock/sheath region, where the speed increase throughout the region indicates that the shock is still driven. 

Region 3 is characterised by a relatively low plasma beta and what appears to be a decreasing speed profile indicating expansion of the magnetic structure. The magnetic field also seems to be somewhat elevated, however, due to the data gap in the first half-period it is not possible to make a reliable estimation about the strength or rotation of the magnetic field. MSL/RAD on the other hand shows a start of another, less rapid decrease at the beginning of this region which might indicate the beginning of the ejecta--related FD. However, the depression does not return to the onset value but is ``interrupted'' by another decrease at the start of region 4. In region 4 we again identify what appears to be stream interface signatures, however, the reliability is questionable due to the MAVEN data gap and moreover MSL/RAD data show a double--decrease structure which indicates that this region might not be a simple stream interface region.

Region 5 is again characterised by an expanding speed profile, the plasma beta is not low at the very start of the region, but drops later. In the second part where the MAVEN data are more consistent, the magnetic field components show a relatively smooth profile, although the magnetic field strength is not really elevated. These are characteristics similar to CME2 identified in ST-A. Finally, in region 6 we observe elevated temperature and speed, which are the typical HSS signatures.

We note that the whole period encompassing \insitu regions 1--6 correspond to a single, long--duration multi--step FD, where different regions (\ie substructures) correspond to different steps of the FD. The unusually large FD amplitude might therefore be related to the complexity of the event, where different steps add--up into a single global FD, as suggested by some previous studies \citep[\eg][]{papaioannou10,dumbovic16}.

\section{Results}
\label{results}	

\subsection{Pre-conditioning of the heliospheric background}
\label{precondition}

 \begin{figure*}
\gridline{\fig{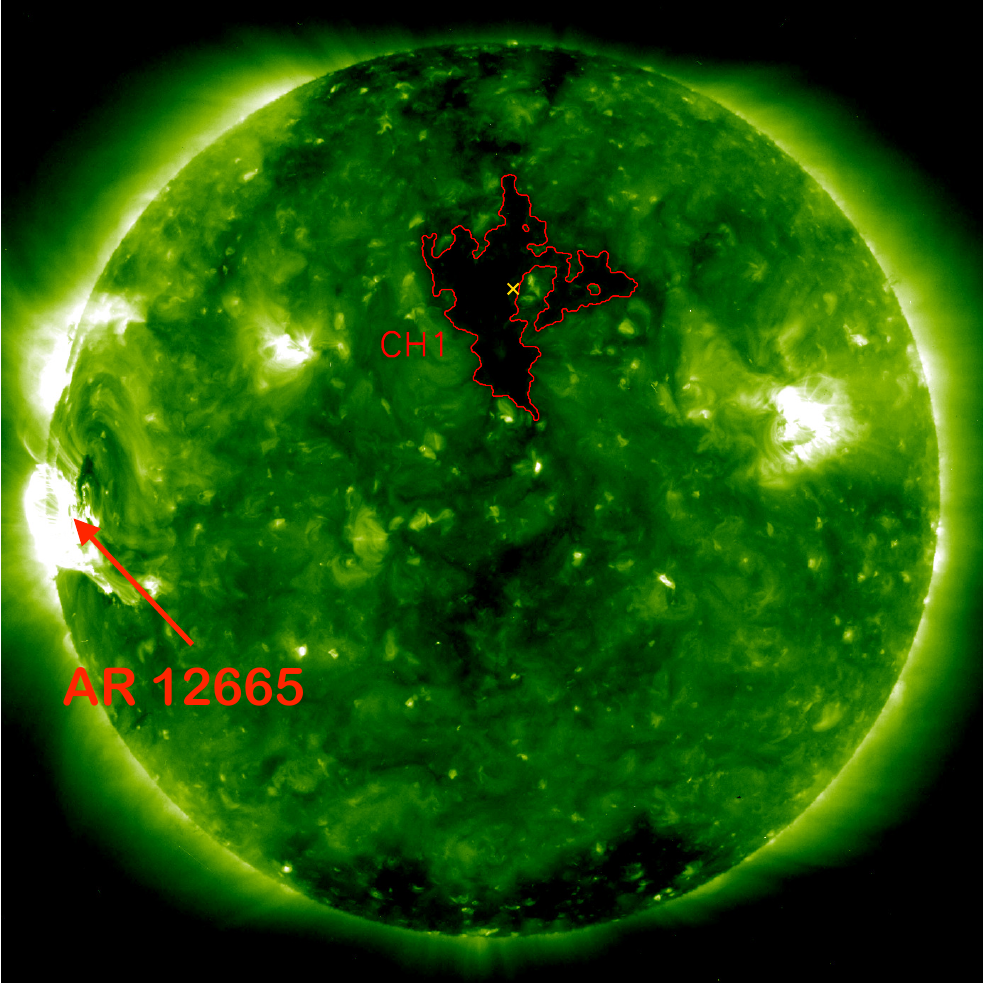}{0.4\textwidth}{(a)}
          \fig{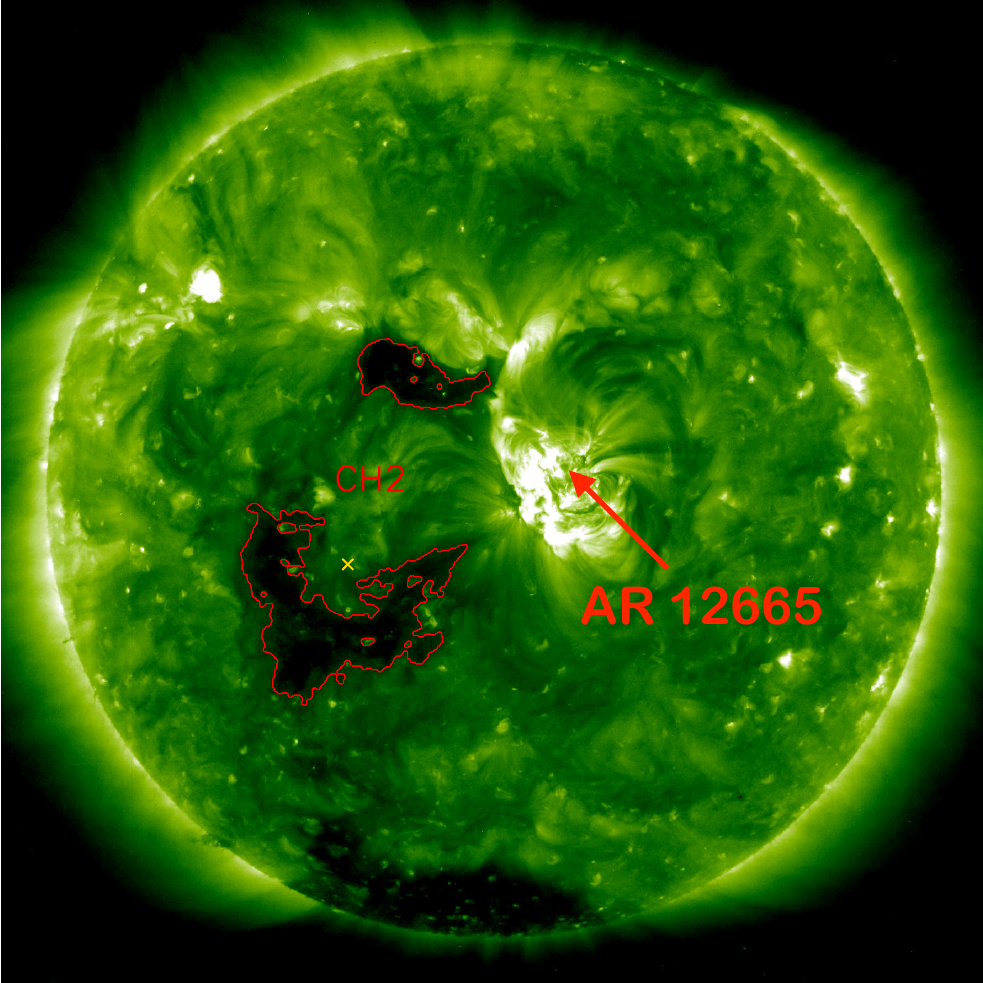}{0.4\textwidth}{(b)}}
\caption{CH1 (a) and CH2 (b) in EUVI/ST-A 195 \angstrom\, crossing the central meridian as observed from ST-A July 23 and July 29, respectively. The red contours mark outline the CHs and the yellow x marks the center of mass of the the CH. The arrow points to AR 12665, which is the source region of CMEs 1 and 2.
\label{fig4}}
\end{figure*}

In order to analyse and understand the propagation and interaction of CMEs 1 and 2 it is imperative to analyse and understand the heliospheric background. For that purpose we check the remote ST-A observations prior to CME 1 and 2 eruptions to identify any relevant coronal holes (CHs) and/or CMEs that might act to pre-condition the interplanetary space. As can be seen in Figure \ref{fig1}d there is a CH passing the central meridian as seen from ST-A on the day of the eruptions (CH1, see Figure \ref{fig4}a). Visually, the CH area crossing the central meridian slice does not appear very large and thus we do not expect the corresponding HSS to be very fast \citep{nolte76,vrsnak07a,tokumaru17,hofmeister18}. Therefore, it might be likely that the two CMEs ``catch-up" with the SIR corresponding to CH1, which might influence their propagation. We observe another possibly relevant CH which is at the time of the eruption behind the eastern limb as viewed from ST-A (CH2, see Figure \ref{fig4}b). CH2 is ``trailing" the eruption site, however due to its vicinity to the active region might interact with the eruptions through interchange reconnection \citep[\eg][]{crooker02}. In Figure \ref{fig4} CH1 and CH2 are shown at the time crossing the central meridian as observed from ST-A, where their corresponding area is extracted using a thresholding technique based on the study by \citet{heinemann18a,heinemann18b}. The CH area values are presented in Table \ref{tab1} for both CHs. CH2 consists of the top and bottom part, therefore we present three area calculations: for the top part only, for the bottom part only and for both (full). In addition, based on the calculated area we calculate the peak speed of the associated solar wind using the CH area -- SW speed relations given by \citet{nolte76} and \citet{tokumaru17}, as well as latitude dependent CH area -- SW speed relation by \citet{hofmeister18}. These are also presented in Table \ref{tab1}. It can be seen that the HSS originating from CH1 is most probably slower than CMEs 1 and 2, hence, a kinematical interaction between them is likely. It can also be seen that CH2 is expected to yield a faster streams than CH1, therefore although the angular separation between the two is roughly $90^{\circ}$, if undisturbed we would expect the time difference between two corresponding SIR to be $<7$ days.
 
\begin{table}
\centering
\caption{CH area values and peak speed of the associated solar wind calculated using the CH area -- SW speed from different studies.}
\label{tab1}
\begin{tabular}{l|c|c|c|c}
							&	CH1				&	CH2 (top)			&	CH2 (bottom)		&	CH2 (full)\\
\hline
$A$[$10^{10}\mathrm{km}^{2}$]	&	$4.8 \pm 0.3$		&	$1.28 \pm 0.05$	&	$5.3 \pm 0.4$		&	$6.5 \pm 0.5$\\
$v$[\kmps] \citep{hofmeister18}		&	$552 \pm 4$		&	$505 \pm 1$		&	$588 \pm 9$		&	$630 \pm 10$\\
$v$[\kmps] \citep{tokumaru17}		&	$476_{-3}^{+5}$	&	$462 \pm 3$		&	$477_{-3}^{+7}$	&	$483_{-3}^{+7}$\\
$v$[\kmps] \citep{nolte76}			&	$790_{-10}^{+60}$	&	$525_{-7}^{+15}$	&	$810_{-20}^{+80}$	&	$910_{-20}^{+90}$\\
\end{tabular}
\end{table}

\begin{figure*}
\gridline{\fig{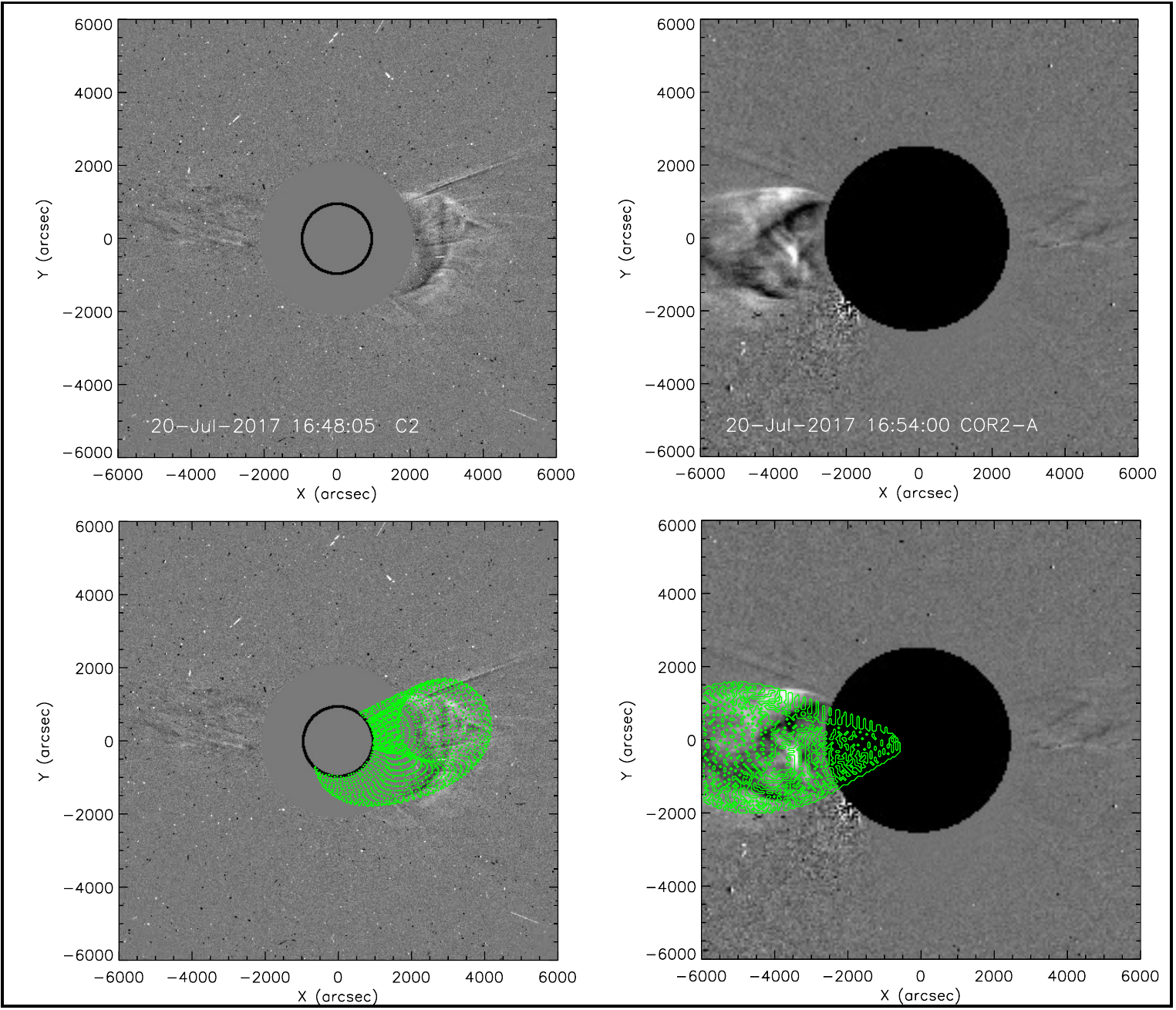}{0.48\textwidth}{(a)}
          \fig{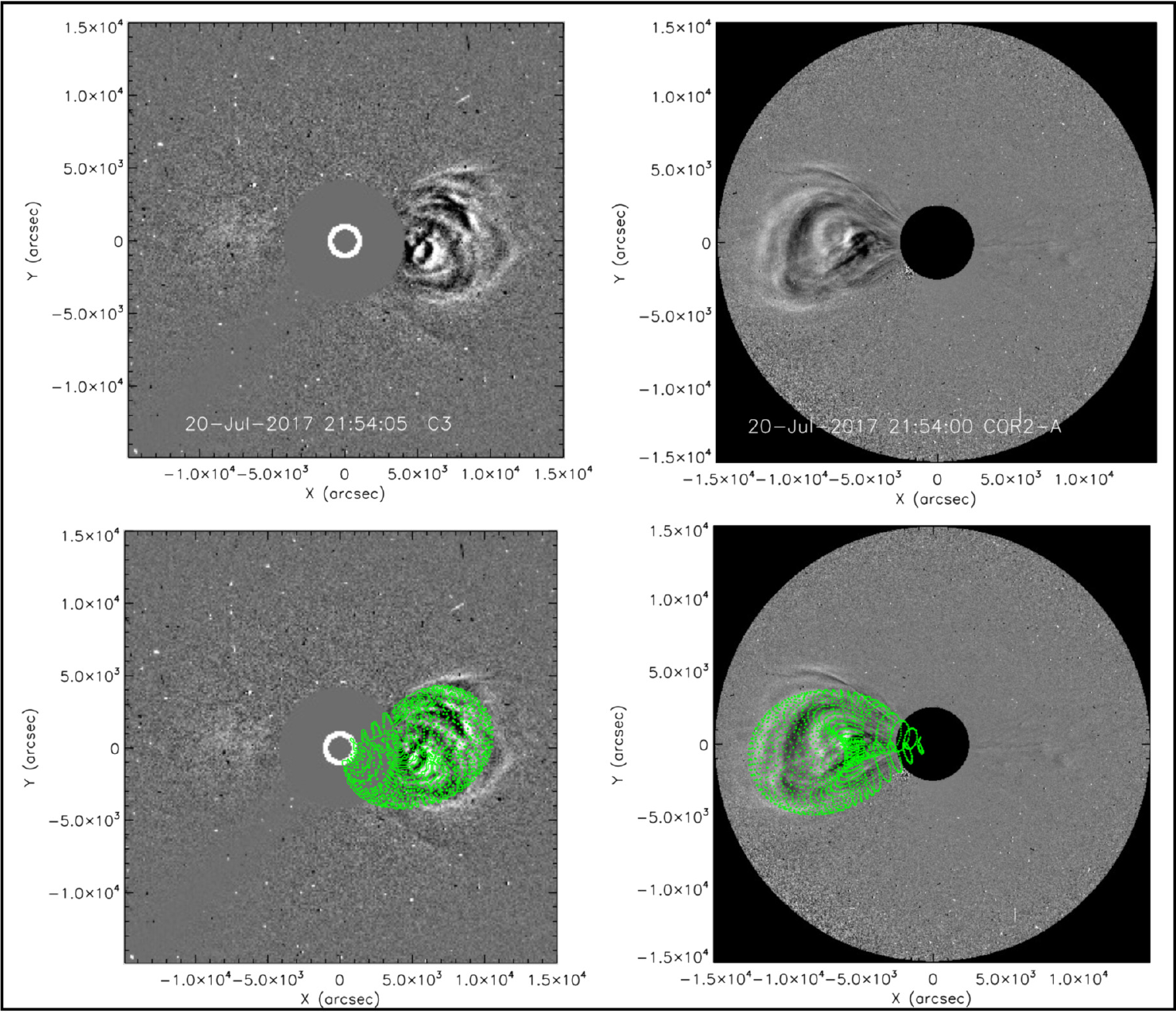}{0.48\textwidth}{(b)}}
\caption{GCS reconstruction of two CMEs that erupted July 20 2017 and pre-condition the interplanetary space:
a) a narrow, faint and slow CME0.1 directed between Mars and STEREO-B;
b) a wider, brighter and faster CME0.2 directed towards STEREO-B.
Parameters of the two CMEs are given in Table \ref{tab2}.
\label{fig5}}
\end{figure*}

In section \ref{data} we noted that in addition to CMEs 1 and 2 we observed two more CMEs that have a position angle relating to Mars direction. These two CMEs (marked as CME0.1 and CME0.2) erupted from the same AR on July 20 2017, and might therefore be important for pre-conditioning of the interplanetary space \citep{temmer17,temmer15}, especially in the direction of Mars. In Figure \ref{fig5} we present the GCS reconstruction of these two CMEs. CME0.1 starts with loop motions observed at the east limb  in the ST-A EUVI 195 \angstrom, followed by the CME observed in COR1 at 13:25 UT. The CME appears in the ST-A/COR2 and LASCO/C2 field of view at 14:35 and 15:48 UT, respectively and appears to be moving linearly with a speed of about 300 \kmps. It is a very faint CME and it is not observed in LASCO/C3. The first signatures of CME0.2 are very similar, with similar loop motions close to the same AR as CME0.1, followed by a CME detection by COR1 at 17:50. The CME appears in the ST-A/COR2 at 19:10 UT, in LASCO/C2 at 18:24 UT, and in LASCO/C3 at 19:30 UT. CME0.2 decelerates in LASCO/C3 and has a highly distorted ``wavy" leading edge, reaching a speed of around 600 \kmps at approximately 20 \Rsun. The interaction of the two CMEs is very likely, but their main propagation direction is oriented more towards STEREO-B. Therefore, we conclude that the interplanetary space will not be affected by the two CMEs in the ST-A direction, but there might be some pre-conditioning in the Mars direction.

\subsection{CME propagation}
\label{propagation}	

In Figure \ref{fig6}a a 2D representation of CME 1 and 2 widths and directions in the HEEQ\footnote{Heliocentric Earth Equatorial} system is shown, as obtained from the GCS reconstruction. It can be seen that if we assume that the direction and the geometry do not change, CME1 is not expected to hit ST-A. Furthermore, CME1 \insitu signatures at ST-A indicate lack of expansion, whereas at Mars there are expansion signatures, indicating CME2 is ``pushing" CME1 much stronger in ST-A direction than in Mars direction, \ie that the CME2 apex is directed closer to ST-A than Mars.  As will be shown in the next section, the propagation/evolution modelling of the CMEs give reasonable results supporting that CMEs 1 and 2 are deflected towards ST-A by 15 and 20 degrees, respectively. We might consider CH2 (see Figure \ref{fig4}a) as a potential deflection source as it lies just east of the AR and deflection by 15-20 degrees would not be an unusual value given its proximity to CMEs source regions \citep[see \eg][]{gopalswamy09}. However, the deflection by CHs usually occurs at lower heights, within a couple solar radii, where we would expect to observe it already in coronagraphs \citep[\eg][]{kay13} and therefore the deflection of CMEs 1 and 2 at heights $>15$\Rsun\, due to CH2 is not likely. Another possible explanation is the deflection in the interplanetary space due to CME-CME interaction \citep{gopalswamy01a}. CME1 might also be deflected due to interaction with CME2 similarly as it would be expected that slow CMEs are deflected towards west when moving in a fast solar wind \citep{wang04,wang14}, however, even considering that the momentum of CME1 is not negligible compared to CME2, this cannot easily explain the deflection of CME2. Finally, one should not forget on the limitations of the observational methods, where a $10^{\circ}$ error in the CME direction is not unusual \citep{mierla10}. We also note that CME2 is a complex CME, which also might influence the determination of the direction. Based on these considerations as well as modelling attempts/efforts, for the purpose of analysis of CME propagation, evolution and their interaction we can assume that the directions of CMEs 1 and 2 in HEEQ system are changed with respect to the ones obtained by the GCS reconstruction (see Table \ref{tab2} and Figure \ref{fig6}c). 

Figure \ref{fig6}b shows the CME 1 and 2 early kinematics, as obtained by the GCS reconstruction. The measured kinematic curves of CME 1 and 2 apexes are given in a distance-time plot, where an error of 5\% was assumed. It can be seen that measured 3D kinematics is in both cases represented well by motion with constant speeds of 950 \kmps and 2700 \kmps, respectively. To estimate the CME speed at the inner boundary for the propagation models used in sections \ref{dbm} and \ref{enlil} (20\Rsun and 21.5\Rsun, respectively) we extrapolate GCS-obtained kinematics assuming constant speed. 

The angular extent of CMEs 1 and 2 in the equatorial plane was estimated as: $\omega_{\mathrm{max}}-(\omega_{\mathrm{max}}-\omega_{\mathrm{min}})\times\mathrm{|tilt|}/90$, where $\omega_{\mathrm{max}}$ and $\omega_{\mathrm{min}}$ are face-on and edge-on widths according to \cite{thernisien11} and the tilt is the angle of the croissant axis with respect to the equatorial plane. This method was used previously to estimate the CME half-width with the assumed cone geometry by \citep{dumbovic18b,guo18b} and takes into account CME tilt, but not the latitude. Therefore, to cross-check the validity of this method we in addition estimate the CME opening angle using the ecliptic cut of the GCS-reconstructions\footnote{this method takes into account both the latitude and the tilt, but the estimation is performed by the observer and is therefore somewhat subjective}, obtaining a similar result (within $\pm5^{\circ}$). For the purpose of simulating an elliptical cross section of the cone, which is needed for numerical simulations shown in Section \ref{enlil}, the method is adapted to calculate the major and minor axis obtaining $r_{\mathrm{min}}=23^{\circ}$ and $r_{\mathrm{max}}=38^{\circ}$ for CME1 and $r_{\mathrm{min}}=22^{\circ}$ and $r_{\mathrm{max}}=40^{\circ}$ for CME2. All initial CME parameters used as an input for propagation models, as well as GCS reconstruction parameters are given in Table \ref{tab2}. We note that GCS reconstruction was performed separately for COR1 and COR2, but for the purpose of heliospheric propagation COR2 reconstruction was used (see section \ref{cme} for details). CMEs 0.1 and 0.2 are included in Table \ref{tab2} as their input was used for simulations in section \ref{enlil}.

\begin{table}
\centering
\caption{CME parameters obtained from GCS reconstruction (a) and used for heliospheric propagation (b)}
\label{tab2}
\begin{tabular}{l|c|c|c|c|c}
	&	CME1	&	CME2 (COR1)	&	CME2 (COR2)	&	CME0.1	&	CME0.2\\
\hline
\multicolumn{6}{l}{a) GCS reconstruction parameters}\\
\hline
longitude (degrees)	&	170	&	185	&	190	&	150	&	117\\
latitude (degrees)	&	10	&	-7	&	-17	&	-5	&	-6\\
tilt (degrees)		&	-5	&	15	&	0	&	-25	&	-37\\
aspect ratio				&	0.35	&	0.5	&	0.9	&	0.27	&	0.35\\
half angle (degrees)	&	25	&	60	&	10	&	20	&	30\\
\hline
\multicolumn{6}{l}{b) initial parameters for heliospheric propagation}\\
\hline
longitude (degrees)			&	-175				&	--	&	-150				&	150				&	117\\
latitude (degrees)			&	10				&	--	&	-10				&	-5				&	-6\\
half angle (degrees)			&	44				&	--	&	74				&	--				&	--\\
$r_{\mathrm{min}}$ (degrees)	&	23				&	--	&	22				&	20				&	23\\
$r_{\mathrm{max}}$ (degrees)	&	38				&	--	&	40				&	32				&	38\\
start date \& time			&	2017-07-23 05:20	&	--	&	2017-07-23 06:00	&	2017-07-21 01:00	&	2017-07-21 02:30\\
speed ($\mathrm{km~s}^{-1}$)	&	950				&	--	&	2700				&	300				&	600\\
\end{tabular}
\end{table}

\subsubsection{Drag-based model (DBM)}
\label{dbm}

We model the kinematics of CMEs 1 and 2 separately for ST-A and Mars directions using the drag-based model \citep[DBM,][]{vrsnak13} adapted for the cone geometry \citep{zic15}. The initial CME parameters are based on the GCS reconstruction (see Table \ref{tab2}), whereas the parameters used to describe the effect of the ambient solar wind are taken to be different in the two directions. In ST-A direction we assume that there is no influence on the CME kinematics by previous 2 CMEs (CME0.1 and 0.2) and therefore assume the ``standard" drag for CME1 \citep[$\gamma=0.2$\gammaunit, see][]{vrsnak13,vrsnak14}. For CME2 we estimate a reduced drag ($\gamma=0.1$\gammaunit) caused by pre-conditioning of the IP space due to CME1. We use a ``standard" solar wind speed for both CMEs in ST-A direction \citep[400 \kmps, see][]{vrsnak14,dumbovic18a}. The solar wind speed observed \insitu before the arrival of the CMEs at ST-A is somewhat higher (500 \kmps), however, we note that the ambient solar wind speed used by DBM corresponds to the solar wind CME moves through, and therefore does not necessarily correspond to the \insitu speed measured before the ICME. On the other hand, the observed solar wind speed after the CME passage, which might be a better proxy for the ambient solar wind speed, is $\sim 400$\kmps.

\begin{figure}[ht!]
\plotone{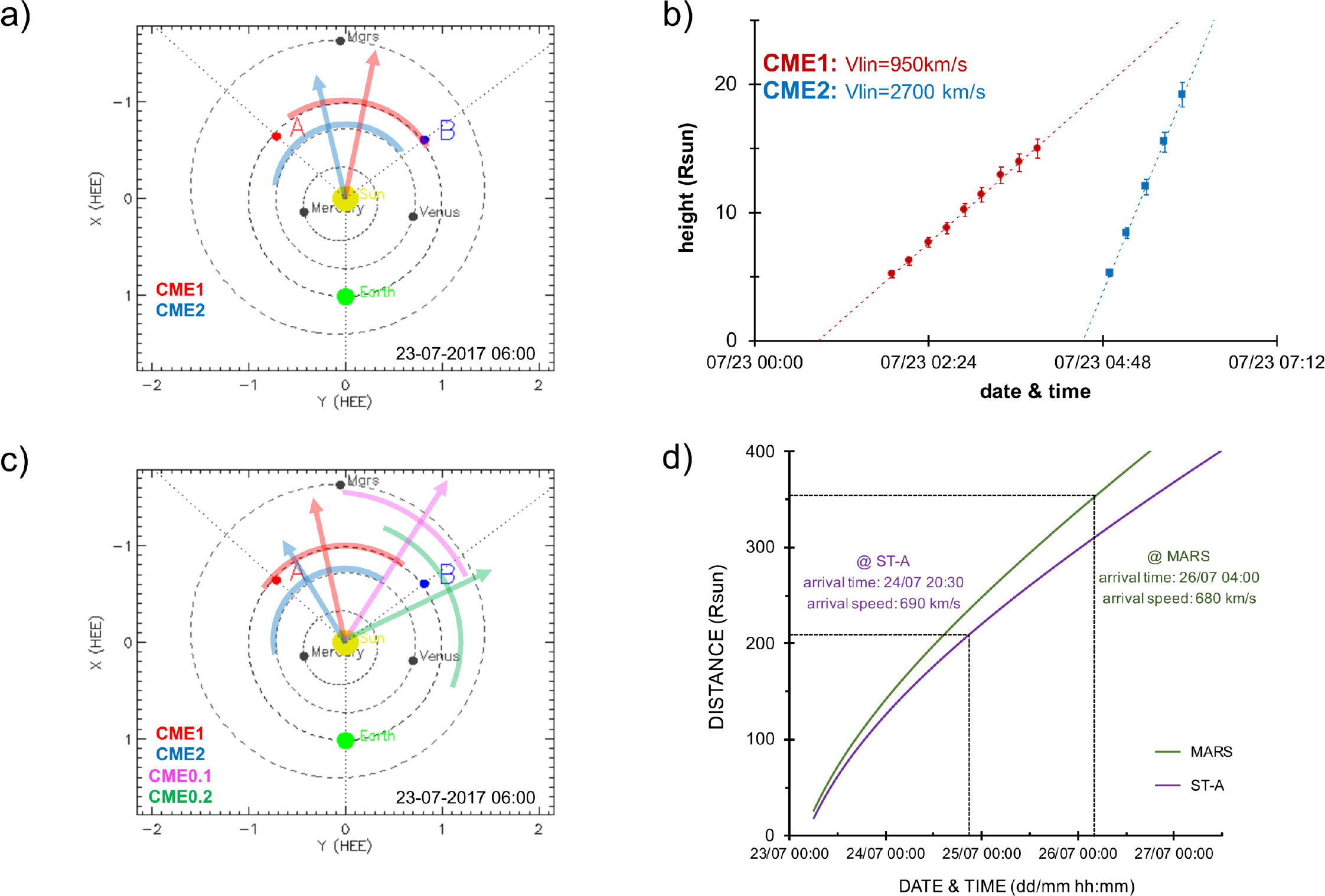}
\caption{CME directions and kinematics obtained by GCS reconstruction and drag-based model (DBM):
a) 2D representation of CMEs 1 and 2 halfwidths and directions in HEEQ system as obtained from GCS reconstruction (the red and blue arrows mark the IP direction of CMEs 1 and 2, respectively, whereas differently coloured arcs provide the angular extent of a corresponding CME);
b) GCS-reconstructed kinematics of the CME1 and CME2 apexes;
c) 2D representation of CMEs 0.1, 0.2, 1 and 2 halfwidths and directions in the HEEQ system as used in modelling (the red, blue, magenta, and green arrows mark the IP direction of CMEs 1, 2, 0.1 and 0.2, respectively, whereas differently coloured arcs provide the angular extent of a corresponding CME);
d) DBM-obtained kinematics of the interacting CMEs (the merged CME1+2 structure) in ST-A and Mars direction.
\label{fig6}}
\end{figure}

In Mars direction we assume that the drag parameter for CME1 is lower than the ``standard" value of $\gamma=0.2$\gammaunit, due to pre-conditioning by previous CMEs (CMEs 0.1 and 0.2) which ``cleared the path", similar as \citet{rollett14}. Based on the \insitu observations by MAVEN and MSL/RAD, but also based on the GCS-related estimates of directions and angular extent of CMEs 0.1 and 0.2, we do not expect their interaction with CMEs 1 and 2 in the Mars direction. However, we find it likely that their propagation partly influences the ambient solar wind through which CMEs 1 and 2 propagate on their way to Mars. Therefore, we estimate a somewhat lowered drag for CME1 ($\gamma=0.15$\gammaunit) and reduced drag for CME2 ($\gamma=0.1$\gammaunit, as assumed for the ST-A direction). In Mars direction we initially use the same solar wind speed as in the ST-A direction, however we note that much better results are obtained with estimated speed of 500 \kmps. We note that solar wind speed observed \insitu before the arrival of the CMEs at Mars is somewhat lower (350 \kmps), on the other hand the observed solar wind speed after the CME passage is $\sim 600$\kmps.

We model CMEs 1 and 2 separately and obtain their DBM kinematical curves for ST-A and Mars direction. For each direction we estimate the interaction point for the 2 CMEs based on the condition that the two kinematical curves cross each other. We note that physically this is highly idealised, because we implicitly assume that the CME is fully represented with an infinitesimally thin leading edge. Nevertheless, the procedure provides a quick and simple estimate of the interaction point, whereas any idealisations can be regarded as introducing an uncertainty in the timing. Similar procedure was used by \cite{guo18b}, where the interaction point was derived based on the kinematical curves for the apexes of the corresponding CMEs assuming that all CME leading edge segments interact at the same time. Here the angular separation of the CME apexes is too large and thus we allow that interaction of different segments can occur at different times. We find that the interaction occurs almost immediately after the CME 2 launch by DBM, at 18.2\Rsun\, and 26\Rsun\, in the ST-A and Mars direction, respectively. 

We assume that the momentum is conserved during the interaction and that after the interaction the 2 CMEs continue to propagate as a merged entity, similar as \citet{temmer12} and \citet{guo18b}. The direction of the merged entity is presumably defined by the direction of the faster CME (which `drives" the whole entity), whereas the width is determined by the angular extent of the wider CME. The mass of the merged entity is given as the sum of masses of the two CMEs, where we assume that both CMEs have comparable masses. We note that this is a somewhat arbitrary mass estimation, because we do not actually measure the CME mass. On the other hand, we note that both CMEs are quite bright and interact very close to the Sun, therefore mass measurements might very likely have large uncertainties and moreover we are only interested in the mass \textit{ratio}. With these assumptions, DBM can be recalculated for the merged entity, where the initial speed is given by the momentum conservation and the new drag parameter is re-calculated as $\gamma_{1+2}=\gamma_{1}M_{1}sin^2\omega_{1+2}/[(M_1+M_2)sin^2\omega_{1}$], where $M_1$ and $M_2$ are masses of CME 1 and 2, $\omega_{1}$ and $\omega_{1+2}$ are half-widths of CME 1 and merged entity, and $\gamma_{1}$ is the drag-parameter of CME 1 \citep[calculation based on equation 2 in][]{vrsnak13}. Note that the re-calculated drag parameter is different in ST-A and Mars direction, as is the solar wind speed. The kinematic curves for the interacting CMEs are plotted in Figure \ref{fig6}d. The estimated arrival time at ST-A is July 24 at 20:30 UT with the arrival speed of 690\kmps, whereas the measured arrival time is July 24 23:00 UT with arrival speed of 660\kmps. The estimated arrival time at Mars is July 26 at 04:00 UT with the arrival speed of 680\kmps, whereas the measured arrival time is July 26 02:00 UT with arrival speed of 700\kmps.

In addition, we perform an ensemble run of the drag-based model \citep[DBEM,][]{dumbovic18a}, where the ensemble input is produced manually based on the standard DBM input as obtained from GCS reconstruction using the method described by \citet{dumbovic18a}. DBEM is not suitable to take into account CME-CME interaction, therefore, we regard only the merged entity as derived by DBM. We produce 48 CME ensemble members separately for ST-A and Mars directions for the merged entity based on the input used in DBM assuming error ranges of 1h for the start time, 200 \kmps for the initial speed and 10$^{\circ}$ for longitude and width, whereas we use 15 synthetic values for solar wind speed and drag parameter assuming error ranges of 0.05 \gammaunit and 50 \kmps, respectively. In ST-A direction DBEM predicts a 100\% arrival probability with likeliest arrival time (median) at July 24 21:30 UT and a spread of $\pm4.5$h, whereas the likeliest arrival speed is 690 \kmps with a spread of $-80$\kmps/$+110$\kmps. In Mars direction DBEM also predicts 100\% arrival probability with likeliest arrival time (median) at July 26 06:00 UT and a spread of $-7.5$h/$+7$h, whereas the likeliest arrival speed is 670 \kmps with a spread of $-60$\kmps/$+80$\kmps. In both cases the observed arrival times and speeds are within the predicted spread. Given the usual forecast errors for CME propagation \citep[\eg][and references therein]{dumbovic18a,wold18,riley18} this is a reasonable agreement with the observations. Therefore, based on the DBM/DBEM results we conclude that it is highly likely that both CMEs interact close to the Sun, arrive ``together" at ST-A and Mars and that the ambient solar wind through which CMEs propagate is different in the ST-A and Mars direction. There are indications that two previous CMEs pre-condition the IP space in the Mars direction lowering the drag, and moreover, that in the Mars direction CMEs propagate through higher-speed solar wind than in ST-A direction.

\subsubsection{WSA-ENLIL+Cone simulation}
\label{enlil}

\begin{figure}
\centerline{\includegraphics[width=0.98\textwidth]{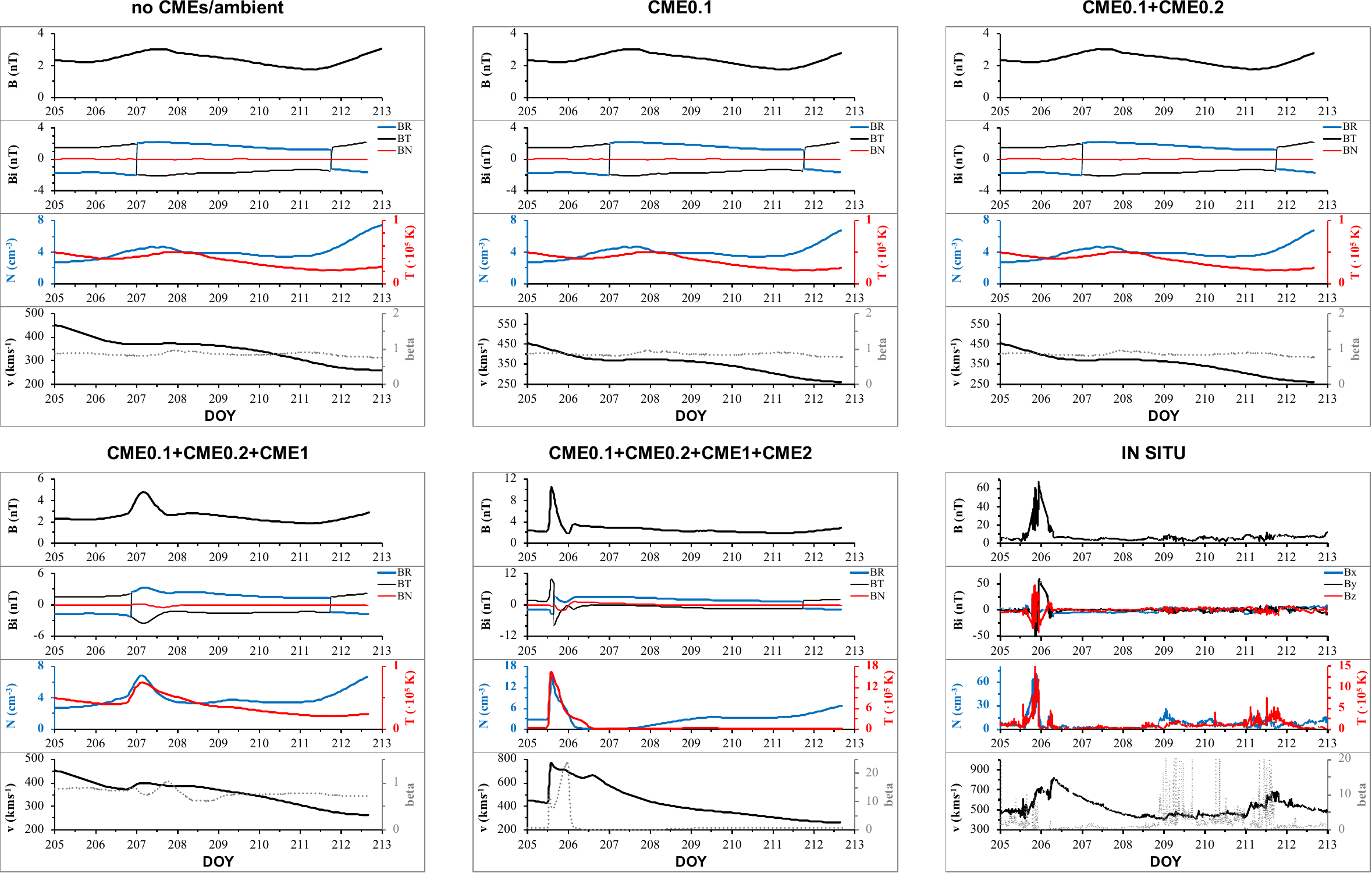}}
\caption{ENLIL multiple block-run results for ST-A for quiet-time ambient (without any CMEs, top left panels), inclusion of CME0.1 (top middle panels), inclusion of CMEs 0.1 and 0.2 (top right panels), inclusion of CMEs 0.1, 0.2 and 1 (bottom left panels), a full suite run with CMEs 0.1, 0.2, 1 and 2 (bottom middle panels) and \insitu measurements by ST-A/PLASTIC (bottom right panels). The panels in each block correspond to (top to bottom) 1-total magnetic field strength, 2-magnetic field strength of three components of the magnetic field, 3-plasma density and temperature, and 4-plasma velocity and beta parameter. The magnetic field components of the ENLIL runs correspond to the RTN coordinates of the HEEQ system, whereas for the ST-A they correspond to RTN coordinates with radial component defined by the Sun-spacecraft line. \textit{(DOY=day of year 2017)}}
\label{fig7}
\end{figure}

We perform a WSA-ENLIL+Cone heliospheric simulation \citep{odstrcil03,odstrcil04,odstrcil05} which is available through the \textit{Runs-on-Request} system of the Community Coordinated Modeling Center (CCMC\footnote{\url{https://ccmc.gsfc.nasa.gov/requests/requests.php}}). The runs consider not only the merging CMEs 1 and 2, but also the pre-conditioning CMEs 0.1 and 0.2 and are based on the initial parameters given in Table \ref{tab2}. They are based on NSO/GONG magnetogram synoptic maps with a boundary condition type of time-independent single daily update. We note that there is an existing run published by \citet{luhman18}, however since the focus of their study are SEPs, they do not consider CMEs 0.1 and 0.2. The input is based on GCS, similar as for DBM, adapted for ENLIL (cone elliptical cross section). The simulation is available at \url{https://ccmc.gsfc.nasa.gov/database_SH/Manuela_Temmer_112618_SH_1.php}.

\begin{figure}
\centerline{\includegraphics[width=0.98\textwidth]{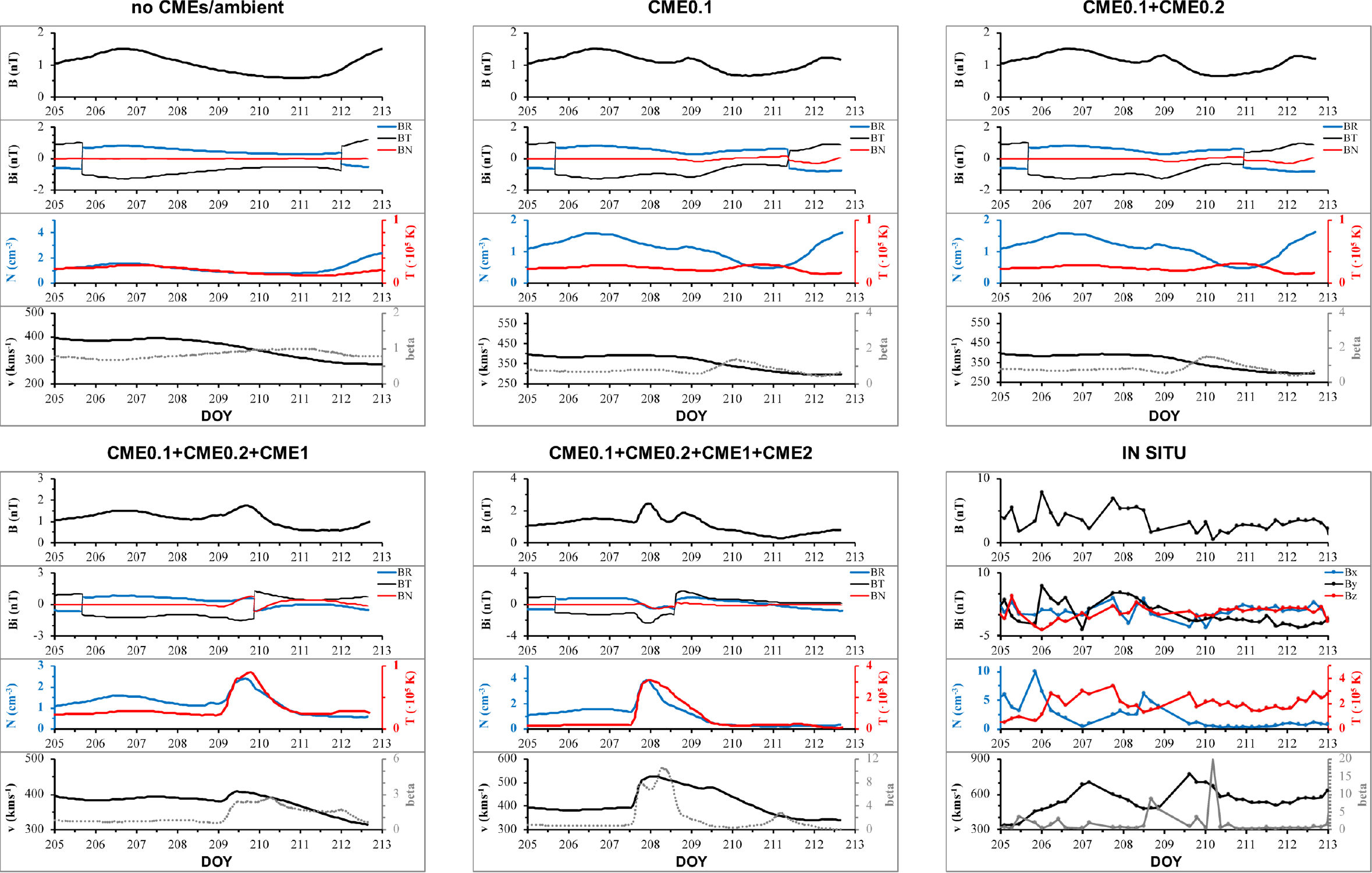}}
\caption{ENLIL multiple block-run results for Mars for quiet-time ambient (without any CMEs, top left panels), inclusion of CME0.1 (top middle panels), inclusion of CMEs 0.1 and 0.2 (top right panels), inclusion of CMEs 0.1, 0.2 and 1 (bottom left panels), a full suite run with CMEs 0.1, 0.2, 1 and 2 (bottom middle panels) and \insitu measurements by MAVEN (bottom right panels). The panels in each block correspond to (top to bottom) 1-total magnetic field strength, 2-magnetic field strength of three components of the magnetic field, 3-plasma density and temperature, and 4-plasma velocity and beta parameter. The magnetic field components of the ENLIL runs correspond to the RTN coordinates of the HEEQ system, whereas for MAVEN they correspond to RTN coordinates of the MSO system, with the radial component defined by the Sun-Mars line.\textit{(DOY=day of year 2017)}}
\label{fig8}
\end{figure}

The simulation reveals that CMEs 0.1 and 0.2 already merge close to the Sun and practically move as one entity, which is a bit slower in the Mars direction compared to STEREO-B and does not arrive at STEREO-A. CMEs 1 and 2 also merge quite close to the Sun and ``pick-up" CMEs 0.1 and 0.2 in the Mars direction. However, it can be seen that the fastest part of the merged CME1+2 entity is directed towards STEREO-A and not Mars, probably due to the longitudinal direction of much faster CME 2 ($150^{\circ}$). The predicted arrival time of the disturbance at ST-A is July 24 around 12:00 UT, which fits very good with observation (July 24 around 13:00 UT, see region 1 in Figure \ref{fig3}-left). The predicted arrival time of the disturbance at Mars is July 26 around 15:00 UT, which is almost 1 day later than the observed shock arrival (July 25 around 17:00 UT, see region 2 in Figure \ref{fig3}-right). In addition, we perform an ENLIL ensemble run \citep{mays15}, where the ensemble input is produced manually based on the standard ENLIL input as obtained from GCS reconstruction using the method described by \citet{dumbovic18a}. The input is slightly different than that for DBEM, because for DBEM we produced a unique ensemble based on the merged entity parameters, whereas for ENLIL we produce ensemble input (24 ensemble members) for each of the 4 CMEs separately. The full input and the results of the run are available at \url{https://iswa.gsfc.nasa.gov/ENSEMBLE/2017-07-20_ncmes4_sims23_HILOX069/}. In ST-A direction the simulation predicts a 100\% arrival probability with median input arrival time at July 24 10:30 UT, average arrival at July 24 11:20 UT and a spread of $-6$h/$+18.5$h. In Mars direction the simulation predicts only 61\% arrival probability with median input arrival time at July 26 06:00 UT, average arrival at July 26 02:00 UT, and a spread of $-14$h/$+6.5$h, \ie the observed arrival time is within the predicted spread. We note that the low hit probability is most likely related to the underestimated longitudinal spread. It is important to highlight that we used CME (magnetic structure) input for the simulations, whereas ENLIL simulates the evolution/propagation of the density disturbance and is therefore more suitable for simulation of the shock. The shock can have a larger spatial extent than the magnetic structure of CME 2, can be driven closer to the Mars direction and not significantly influenced by July 20 CMEs. However, in our study we are more interested in the qualitative description of the CME 1 and 2 interaction, than accurate CME 2 shock propagation, therefore we do not attempt to obtain a better arrival time match by introducing the shock-related input.

For a better qualitative analysis of the events observed at ST-A and Mars we use the ENLIL multiple block runs. The multiple block runs allow us to see how the simulation results change by consecutively adding CMEs in the simulation. The first run describes the ambient medium, without any CMEs, the second run includes the first CME of July 20 (marked as CME0.1), the third run includes both the first (CME0.1), as well as the second CME of July 20 (marked CME0.2), the fourth run includes both CMEs of July 20 and CME1, and finally the last run includes all four CMEs. The performed multiple block runs are available as separate synthetic \insitu profiles at different targets in the output files of the simulation. In Figures \ref{fig7} and \ref{fig8} ENLIL multiple block run results are shown for heliospheric positions corresponding to ST-A and Mars, respectively. From top-left to bottom-right different panels show runs starting from no CMEs to inclusion of all 4 CMEs (as described above), where the last panel shows \insitu measurements on the same time scale. 

 \begin{figure}
\centerline{\includegraphics[width=0.9\textwidth]{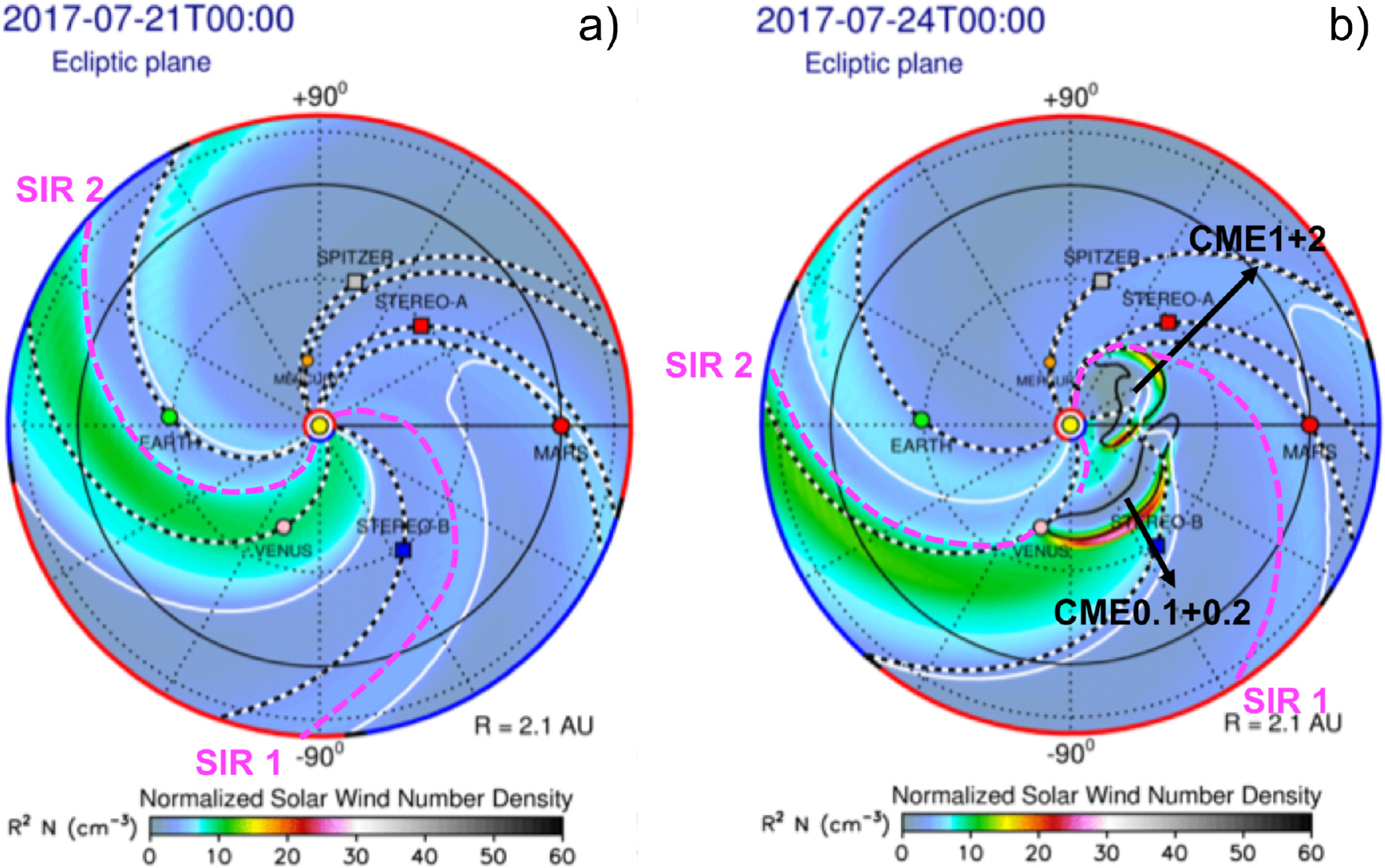}}
\caption{Selected timeframes of the ENLIL simulation: quiet time (a) and propagation phase (b), where first propagating disturbance represents merged CME0.1 and 0.2 entity (CME0.1+0.2) and the second merged CME 1 and 2 entity (CME1+2). SIR1 and SIR2 are approximatively highlighted at the density change by a magenta dashed line. SIR2 is disconnected before and after the CMEs in (b).}
\label{fig9}
\end{figure}

Two sector boundaries are observed in the multiple block-run plots in Figure \ref{fig7}. The first one corresponds to the SIR originating from the CH observed to pass the central meridian (as seen from Earth) around July 5 and very likely corresponds to CH1 analysed in Section \ref{precondition}. The second one corresponds to the SIR originating from the CH eastward of the AR from which CME 1 and 2 originate, which is observed to pass the central meridian (as seen from Earth) around July 13 and very likely corresponds to the CH2 analysed in Section \ref{precondition}. It can be seen that inclusion of 2 CMEs that preceded CMEs 1 and 2 do not impact the \insitu observations at ST-A. Furthermore, it can be seen that in a full suite run (CME0.1+CME0.2+CME1+CME2) there is a qualitative agreement with the observed \insitu double-CME signatures. Moreover, the simulation indicates that SIR1 is ``squeezed" in the sheath region of CMEs 1 and 2. Although this cannot be confirmed by \insitu measurements due to the data gap, we note that we do not find SIR1 signatures in front of CMEs 1 and 2. We also note that the complete lack of an expanding profile of CME1 also indicates there might be a structure in front of it, inhibiting its expansion, with SIR1 being the only candidate to explain this. Therefore, we argue that it is indeed possible that the SIR1 is ``squeezed" in the sheath region of CMEs 1 and 2, as the simulation indicates. The timing of the second sector boundary roughly corresponds to the observed SIR signatures between DOY 211 and 212, likely originating from CH2. 

In Figure \ref{fig8} again two sector boundaries can be observed in the ENLIL multiple block runs, corresponding to the same CH as the ones observed in Figure \ref{fig7}. However, it can be seen that in the Mars direction the first SIR reaches Mars before it interacts with CMEs 1 and 2 and roughly corresponds with the timing of the SIR signatures observed in region 1 in Figure \ref{fig3}. On the other hand, according to the simulation SIR2 is strongly affected by CMEs. It can be seen that inclusion of CME0.1 and CME0.2 impacts \insitu observations at Mars, but their signatures are entirely lost in a full suite run. This indicates that CME0.1 and CME0.2 affect the ambient solar wind, but we would not expect to see their significant impact at Mars. Furthermore, in a full suite run we observe SIR2 being ``squeezed" between CMEs 1 and 2, which seems to be in contradiction with the fact that CH2 is trailing the AR and that the two CMEs merge already in the corona. While this might be an artefact related to the fact that ENLIL runs do not involve magnetic structures (CMEs are introduced as pressure pulses), it is important to note that the SIR signatures are indeed observed in the MAVEN data between the two CMEs, with the timing roughly corresponding to SIR2 from simulations. A possible explanation is that an interchange reconnection occurs between the open field lines of CH2 and the leg of CME1 \citep[similar as described by][]{crooker02}. This might result in open field lines of both CME1 and the nearby CH2 (see the northern part of CH2 in Figure \ref{fig4}b) being caught between CMEs 1 and 2 producing SIR-like signatures between the two CMEs. This would lead to SIR-like signatures observed at Mars, but also at ST-A, where we indeed do find SIR signatures. However, the latter is not expected from the ENLIL simulation. 

We want to emphasise that the quantitative correspondence between synthetic \insitu measurements of the ENLIL multiple block runs and real \insitu measurements by MAVEN and ST-A is quite poor (\eg\, magnetic field at ST-A is 4 times weaker in the simulation). On the other hand, we find that the multiple block runs in synergy with the \insitu measurements are very insightful in understanding the ``chronological" order of sub-structures observed \insitu.

\subsection{CME evolutionary properties}
\label{evolution}	

In order to analyse how CME1 evolves, \ie how the size and the magnetic field change with heliospheric distance, we compare the initial FR properties (obtained from GCS) with \insitu-measured properties at ST-A and Mars. For that purpose, we estimate the initial radius of the flux rope based on the last time step of the performed GCS reconstruction, which is taken as the initial time of the FR evolution. The radius of the FR obtained from the GCS reconstruction is different across the FR - it is largest at the apex and smallest at the flanks. Based on the direction of CME1 and the relative positions of the spacecraft, as the starting radius in the Mars direction we take the apex radius and in the ST-A direction we take the flank radius and we assume a 5\% error (see Table \ref{tab3}). Based on the start time and the observed \insitu arrival time we calculate the transit time to ST-A and Mars and assume $\pm1$h error. The final radius of the FR at ST-A was obtained based on the FR orientation and impact parameter obtained from the Lundquist fitting \citep[][]{lundquist51,leitner07}, according to the method described by \citet{vrsnak19}. At Mars we estimate the radius assuming that the spacecraft passes the FR vertically with respect to the FR axis using the expanding profile of plasma flow speed. The error of the radius is determined \textit{via} error propagation assuming SW speed and FR duration errors of 5\%. The central axial field strength at ST-A is obtained from the Lundquist fitting, whereas at Mars we estimate it based on \insitu measurements assuming it corresponds to the maximum measured value (note that there is a data gap at the start of the ICME, therefore, the maximum is around ICME center). In both cases we assume an error of 1 nT, which is around 15\% for Mars, but we note that due to the MAVEN data gap at the front of the CME it is reasonable to assume larger error. Using these \insitu values we calculate the axial magnetic flux of the flux rope using equation 52 from \citet{devore00} derived for Lundquist-solution FR, $\Phi_{\mathrm{FIN}}=1.4B_{\mathrm{c,FIN}}a^2_{\mathrm{FIN}}$, where $B_c$ is the central axial field strength. The obtained flux values at Mars and ST-A are different (see Table \ref{tab3}), indicating different evolution in two different directions of propagation. Note that the result shown in Table \ref{tab3} is closely related to the assumption of the Lundquist solution, whereas considering different  sets of assumptions might yield that the factor of 2 difference in the flux is not significant.

Assuming that the FR radius $a$ is expanding self-similarly following a power-law, $a=a_0(R/R_0)^{n_a}$ \citep[see \eg][]{demoulin08,gulisano12,vrsnak19} the power-law index $n_a$ can be derived using the initial and final distance/FR radius. We note that $n_a=1$ indicates an isotropic self-similar expansion, whereas $n_a<1$ and $n_a>1$ could indicate  a weaker or stronger expansion, respectively, as the observational studies roughly constrain $n_a$ to $0.45<n_a<1.14$ \citep[\eg][]{bothmer98,leitner07,gulisano12,vrsnak19}. Using the values presented in Table \ref{tab3} we calculate the power-law index $n_a$ in ST-A and Mars direction separately, allowing that the FR expands differently in two different directions. The errors are calculated by estimating the upper and lower power-law curves based on the error bars of the datapoints \citep[for a more detailed description see][]{vrsnak19}. We observe that $n_a$ is much smaller in ST-A direction than Mars direction, indicating that radial expansion is much weaker in the ST-A direction, in agreement with \insitu observation.

\begin{table}
\centering
\caption{Flux rope initial (GCS), \insitu and evolutionary characteristics}
\label{tab3}
\begin{tabular}{l|c|c}
												&	ST-A									&	Mars\\
\hline
start time											&	2017-07-23 03:54 UT					&	2017-07-23 03:54 UT\\
start height, $R_0$									&	15 \Rsun								&	15 \Rsun\\
GCS radius, $a_0$									&	$(2.8\pm0.1)$ \Rsun						&	$(3.9\pm0.2)$ \Rsun\\
\hline
final distance, $R(t)$									&	$208$\Rsun							&	$352$\Rsun\\
transit time										&	$(43\pm1)$ h							&	$(70\pm1)$ h\\
\insitu radius, $a_{\mathrm{FIN}}$						&	$(13\pm2)$ \Rsun						&	$(48\pm5)$ \Rsun\\
\insitu central magnetic field strength, $B_{\mathrm{c,FIN}}$	&	$(40\pm1)$ nT							&	$(7\pm1)$ nT\\
\insitu magnetic flux, $\Phi_{\mathrm{FIN}}$				&	$(5\pm1)\cdot10^{20}$ Mx				&	$(1.1\pm0.3)\cdot10^{21}$ Mx\\
\hline
power-law radius expansion index, $n_a$					&	$0.58^{+0.07}_{-0.08}$					&	$0.80\pm0.05$\\
FD magnitude										&	--									&	$(2.8\pm0.2)\%$\\
power-law magnetic field expansion index, $n_B$			&	$\sim1.8$								&	$\sim2.1$\\
expansion type, $x=n_B-2n_a$							&	$x>0$								&	$x>0$\\
initial central magnetic field strength, $B_0$				&	$\sim0.05$G							&	$\sim0.05$G\\
initial axial flux, $\Phi_0$								&	$\sim5\cdot10^{21}$ Mx					&	$\sim5\cdot10^{21}$ Mx\\
\hline
\end{tabular}
\tablecomments{$B_0$ and  $\Phi_0$ were modelled based on cosmic ray data and are assumed to be the same for both directions (for a detailed explanation on the calculation see main text).}
\end{table}

Self-similar expansion of the FR is also related to the drop in the magnetic field, which can also be described \textit{via} power-law, $B=B_0(R/R_0)^{-n_B}$ \citep[see \eg][]{demoulin08,gulisano12,vrsnak19}. We note that $n_B=2$ indicates an isotropic self-similar expansion, but observations constrain it to $0.84<n_B<2.19$ \citep[\eg][and references therein]{gulisano12,vrsnak19}. Deriving the power-law index $n_B$ is not trivial, because we do not know the initial central magnetic field strength. For that purpose, we use the \textit{ForbMod} model \citep{dumbovic18b} which describes interaction of galactic cosmic rays (GCRs) and a flux rope during its propagation and evolution in the heliosphere. The model assumes that the FR is initially empty of GCRs, that it expands self-similarly and fills-up slowly with GCRs which diffuse into it. As a result, after a certain time the FR will be partly filled with particles and consequently GCR observation will show a drop during the FR passage, \ie a Forbush decrease. The FD amplitude at a specific heliospheric distance therefore depends on the expansion rate of the FR and the diffusion rate (\ie diffusion coefficient) and is given by an analytical expression \citep[for a detailed description of the model see][]{dumbovic18b}. Using the measured FD amplitude and with several simple assumptions we can use the model to estimate the initial central magnetic field strength. Firstly, we assume that the observed FD amplitude can be associated with particles of a relatively narrow specific energy range, \ie that MSL/RAD is mostly sensitive to particles of certain energy. We note that this a somewhat arbitrary assumption, because MSL/RAD measurements have contributions from different particles, \ie radiation sources, including both primary and secondary GCRs, electrons, etc. Nevertheless, MSL/RAD measurements were found to be a very useful proxy for GCRs, especially regarding FD measurements \citep{guo18}. The Martian atmosphere shields away lower-energy GCRs ($<150$ MeV, \ie 0.55 GV) and modifies incoming GCR spectra so that higher-energy particles are less modulated \citep{guo18}. On the other hand, the interplanetary GCR spectrum drops very fast with rigidities above 1 GV, thus it is reasonable to assume that MSL/RAD is mostly sensitive to particles around a certain energy. We use a particle transport model, the Atmospheric Radiation Interaction Simulator \citep[AtRIS,][]{banjac19}, which was validated for MSL/RAD by \citet{guo19}, and estimate that MSL/RAD is mostly sensitive to particles of rigidity $\sim2.5$GV. Secondly, we assume that the expression for the diffusion coefficient is governed by the power-law behaviour due to its relation to the magnetic field (\ie $D\sim1/B$) and that it is scaled to the radial perpendicular diffusion coefficient at Earth as given in \citet{potgieter13}. With these assumptions and fitting the model to observations (see Figure \ref{fig3}-right) it is possible to estimate $n_B$ in the Mars direction \citep[see equation 1 in][]{rodari19} and therefore make an order of magnitude estimation of the initial FR central magnetic field strength and flux. 

We find that the magnetic field inside the FR drops faster in the Mars direction than ST-A direction, which is in agreement with \insitu observations, where the magnetic field at ST-A is unusually strong compared to MAVEN. Namely, if we calculate the power-law index of the magnetic field drop, $B=B_0(R/R_0)^{-n_B}$, using $B_0$, $R_0$ and $B$, $R$ for ST-A and Mars, respectively, $n_B=3.3$ is obtained, which is much higher than observationally constrained values, indicating that the magnetic field drop rate is in fact different in the ST-A and Mars directions, as obtained from the \textit{ForbMod} calculation.

Based on the self-similar expansion expressions and considering a Lundquist-type solution with a circular cross-section, it can be seen that the axial magnetic flux rope can also be expressed by a power-law as $\Phi=\Phi_0(R/R_0)^{-x}$, where $x=n_B-2n_a$ denotes an expansion type \citep[for $x=0$ the flux is conserved, for $x<0$ increased for $x>0$ decreased, see][]{dumbovic18b}. Based on these considerations we find that both in the ST-A and Mars direction $x>0$ indicating that the axial flux of the FR is effectively reduced, possibly through erosion due to interaction. Assuming that the initial flux is the same in both directions, our results suggest that the axial flux might be reduced more efficiently in the ST-A direction than Mars direction (Table \ref{tab3}), however, we note that the uncertainties are too large to make a more reliable conclusion.

\section{Discussion and conclusion}
\label{discussion}

We study the solar and interplanetary sources of one of the biggest FDs observed at Mars on July 24 2017. The analysis of \insitu observations at ST-A and Mars presented in Sections \ref{sta} and \ref{mars} indicates that regions 3/yellow and 5/red in Figure \ref{fig3} correspond to two ICMEs, most likely CME1 and 2. Both at ST-A and Mars shock/sheath signatures are observed, however in ST-A direction there is a data gap in the shock/sheath region which hampers a reliable and uniqe interpretation of the \insitu signatures within this region. Both at ST-A and Mars an interaction region is observed between the two CMEs (region 4/orange in Figure \ref{fig3}), however the interaction region is much longer/broader at Mars compared to ST-A. Finally, at Mars a HSS signature is observed at the back of the second ICME (region 6/green in Figure\ref{fig3}), whereas such region is not observed in ST-A. Combining the DBM results with the \insitu signatures described in Sections \ref{sta} and \ref{mars} there are strong indications that regions 3/yellow and 5/red in Figure \ref{fig3} correspond to CME1 and CME2 at ST-A and Mars, respectively. In addition, we do not observe \insitu signatures of CMEs 0.1 or 0.2 at ST-A or Mars. Finally, we note that region 6 in Figure \ref{fig3} (with high-speed stream signatures) supports the DBM result of CMEs propagating through higher-speed solar wind in the Mars direction.

The ENLIL simulations are in agreement with the DBM results and \insitu interpretations of CMEs 1 and 2 corresponding to regions 3 and 5 at ST-A and Mars, respectively. Moreover, the simulations indicate that SIR1 is ``squeezed" in front of CME1 in the ST-A direction, which might be in agreement with a very intense second part of the shock/sheath (region 2/blue) observed in Figure \ref{fig3}. However, due to the data gap in the \insitu measurements this cannot be confirmed. On the other hand, in Mars direction SIR1 arrives shortly before CME1, corresponding to the stream interface signatures and a first decrease in RAD data observed in region 1/green in Figure \ref{fig3}.

The most likely source of SIR1 is a CH passing the central meridian of ST-A at the day of eruption (CH1, see Figure \ref{fig4}a). The simulation is also in agreement with the expansion profiles observed \insitu at ST-A and Mars. In the ST-A direction CME1 is, according to the simulation, constrained between CME2 and SIR1 and thus should not expand notably, as is indeed observed. In the Mars direction on the other hand, both the simulations and \insitu measurements indicate that SIR1 is not ``pushed" by CME1. Therefore, we might expect CME1 to expand more freely in the Mars direction, in agreement with the expanding profile observed in the \insitu measurements. The analysis of the CME evolutionary properties further support the interpretation that the expansion of CME1 is hindered in ST-A direction, whereas it expands more freely in the Mars direction.

The most likely source of SIR2 is a horseshoe-shaped CH lying just next to the active region where CMEs 1 and 2 originate (CH2, see Figure \ref{fig4}b). The simulation shows how CMEs 1 and 2 ``disconnect" SIR2 which partly becomes trapped between CMEs 1 and 2, whereas the HSS continues to stream out from CH2, propagating behind the two interacting CMEs in the ST-A direction. These simulation results are probably related to the fact that the simulations do not involve magnetic structures, on the other hand they do agree with the \insitu measurements showing SIR signatures between two CMEs both at ST-A and Mars. These specific CME-SIR-CME signatures could be related to the interchange reconnection \citep[\eg][]{crooker02} of the CME1 leg with a a nearby CH2, where the open field lines get ``squeezed" between CMEs 1 and 2, whereas the majority of the HSS originating from CH2 propagates/corotates behind the CME2, lagging significantly behind CMEs in ST-A direction, but not in the Mars direction.

Finally, we might consider the possible consequences such an extreme space weather event might have had if it was Earth-oriented. In Figure \ref{fig3} a very strong south-oriented $B_z$ can be seen at ST-A, which could have produced a serious geomagnetic storm at Earth. \citet{liu19} calculated it would cause a severe geomagnetic storm with a Disturbance storm time index $Dst<-300$nT. We note that \citet{liu19} offered an alternative interpretation of the solar sources of the IP signatures observed by ST-A and Mars. They, however, also stress the complexity of the event, as well as preconditioning.

To summarise: based on the presented analysis which combines multi-instrument and multi-spacecraft measurements, as well as different modelling approaches, we find that peculiar \insitu signatures at ST-A and Mars can be explained by CME-CME and CME-SIR interactions. In ST-A direction the interaction inhibited the expansion of the first CME thus resulting in magnetic cloud signatures with an extremely high magnetic field strength and of extremely short-duration. On the other hand, in the Mars direction CME-CME interaction accompanied by interaction with the ambient interplanetary plasma resulted in a complex, long-duration IP disturbance with many substructures, each of which added to the multi-step Forbush decrease, thus producing one of the biggest FDs ever detected on Mars. We underline that there is a highly speculative aspect of the proposed scenario as it is based on the methods and models suffering from significant uncertainties. Hence, other scenarios explaining this complex event might also be considered. However, we note that despite the uncertainties, different methods and models used in this study overlap in the proposed scenario, making it the most plausible explanation of the complex events.

\acknowledgments

The research leading to these results has received funding from the European Union's Horizon 2020 research and innovation programme under the Marie Sklodowska-Curie grant agreement No 745782. J. G. is partly supported by the Key Research Program of the Chinese Academy of Sciences under grant no. XDPB11. S.G.H. and M.T. acknowledge funding by the Austrian Space Applications Programme of the Austrian Research Promotion Agency FFG (ASAP-13 859729, SWAMI). S.H. thanks the OEAD for supporting this research by a Mariett-Blau-fellowship. J. H. acknowledges the support from the MAVEN project. C.M. thanks the Austrian Science Fund (FWF): P31521-N27. T.A. and J.H. thank the Austrian Science Fund (FWF): P31265-N27.
A.M.V., and K.D. acknowledge funding by the Austrian Space Applications Programme of the Austrian Research Promotion Agency FFG (ASAP-14 865972 - SSCME).
The WSA-ENLIL+Cone simulation results were provided by the CCMC through their public Runs-on-Request system and the ensemble results by special request (\url{http://ccmc.gsfc.nasa.gov}; run numbers Mateja\_Dumbovic\_010518\_SH\_1, Manuela\_Temmer\_112618\_SH\_1, 2017-07-20\_ncmes4\_sims23\_HILOX069). The WSA model was developed by N. Arge (NASA GSFC), and the ENLIL Model was developed by D. Odstrcil (GMU). The MSL and MAVEN data used in this paper are archived in the NASA Planetary Data System's Planetary Plasma Interactions Node at the University of California, Los Angeles. The PPI node is hosted at \url{https://pds-ppi.igpp.ucla.edu/}. The usage of the MAVEN/MAG data has been consulted with the instrument PI John Connerney. MSL/RAD is supported in the United States by the National Aeronautics and Space Administration's Human Exploration and Operations Mission Directorate, under Jet Propulsion Laboratory subcontract 1273039 to Southwest Research Institute, and in Germany by the German Aerospace Center (DLR) and DLR's Space Administration Grants 50QM0501, 50QM1201, and 50QM1701 to the Christian Albrechts University, Kiel.

\clearpage

\bibliographystyle{plainnat}
\bibliography{REFs}

\end{document}